\documentclass[aps,prx,twocolumn,superscriptaddress,english,floatfix,longbibliography]{revtex4-2}

\usepackage{graphicx}
\usepackage{physics}
\usepackage{cancel}
\usepackage{enumerate}
\usepackage{dsfont}
\usepackage[colorlinks,citecolor=blue,linkcolor=blue,urlcolor=blue,filecolor=black]{hyperref}
\usepackage{textcomp}
\usepackage{amsmath}
\usepackage{amssymb}
\usepackage{soul}
\usepackage[normalem]{ulem}
\usepackage{lipsum}
\usepackage{mathrsfs}
\usepackage[english]{babel}
\usepackage{bm}
\usepackage{comment}
\usepackage{xcolor}
\definecolor{danielcolor}{RGB}{0, 0, 200}    
\newcommand{\dk}[1]{\textcolor{danielcolor}{#1}}

\begin{document}

\title{Flux noise without flux tunability in superconducting qubits}

\author{Daniel Kruti}
\author{Roman-Pascal Riwar}
\affiliation{Peter Grünberg Institute (PGI-2), Forschungszentrum Jülich, 52425 Jülich, Germany}

\date{\today}

\begin{abstract}
Flux noise is unanimously recognised as a leading dephasing mechanism for flux-tunable superconducting qubits. However, our microscopic understanding remains incomplete, and basic effects like Faraday’s law of induction have only very recently come into focus. Based on a quantum geometric description of the Faraday effect, we provide an in-depth derivation of the coupling of generic magnetic sources to thin film superconducting structures, under appropriate consideration of the device geometry. We apply the resulting framework to time-varying magnetic dipoles, describing surface or substrate spins, as well as current-carrying flux lines. We show that flux noise not only affects dephasing, but also provides a fundamental limit for the qubit quality factor --- notably, even when the qubit contains no loops and is thus nominally \textit{not flux-tunable}. Assuming surface spins as the origin for universal flux noise, we expect that this quality factor limit might be reached in the near term. For flux lines, we formulate a minimal safety distance to conserve the qubit performance, potentially constraining the scale-up of quantum hardware. This distance is boosted in the presence of large capacitor wings typical for transmons, due to a lensing of the electromotive field which is largely independent of Meissner screening.
\end{abstract}

\maketitle

\section{Introduction}

Experimental data from flux-sensitive superconducting circuits, collected over many decades~\cite{Koch_1983,Wellstood_1987,Yoshihara_2006,Kakuyanagi_2007,Bialczak_2007,Sendelbach_2008,Sendelbach_2009,Anton_2012,Anton_2013_exp,Bylander_2011,Gustavsson_2011,Sank_2012,Fink_2013,Yan_2016,Kumar_2016,Quintana_2017,de_Graaf_2017,Rower_2023,Gao_2025}, established a flux power noise spectrum of the form $S_\Phi(\omega)\sim A^2/\omega^\eta$ with $\eta$ close to (slightly below) $1$, and the amplitude $A$ ranging between $1\sim 10 \mu\Phi_0$ ($\Phi_0$ is the flux quantum). This simple power law was reported to remain valid from $10^{-5} \text{Hz}$ all the way up to the low $\text{GHz}$ regime~\cite{Quintana_2017}, with possible deviations when applying in-plane magnetic fields~\cite{Rower_2023}. While its microscopic origin remains unresolved to this day, clusters of surface spins are theoretically considered one of the likeliest candidates~\cite{Koch_2007,Bialczak_2007,Faoro_2008,Anton_2013,Lanting_2014,LaForest_2015,Aquino_2022}. 

\begin{figure*}
    \centering
    \includegraphics[width=1\linewidth]{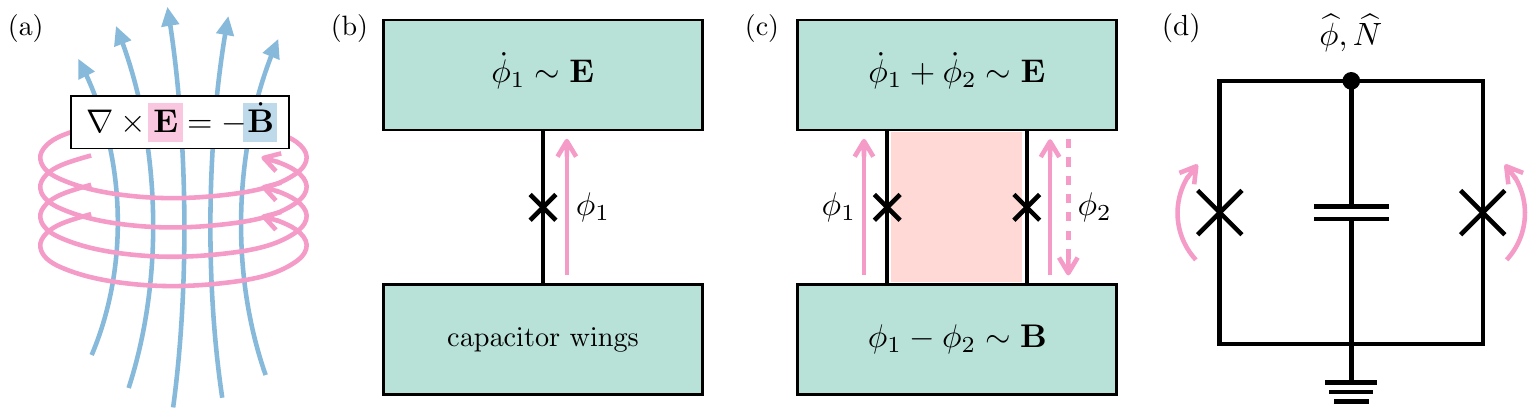}
    \caption{Main concept developed in this work, based on Faraday's law of induction. A time-dependent or fluctuating magnetic field $\mathbf{B}$ induces a corresponding electromotive force (EMF) via the electric field $\mathbf{E}$ (a). Flux noise couples to single-junction (b) as well as double-junction (c) transmon qubits (with capacitor wings) and thus affects qubit performance (in particular the quality factor) independent of flux-tunability. Assigning to each junction $\nu$ a Peierls-like phase drop $\phi_\nu$, we can distinguish between the regular flux mode and an EMF mode. The former enters via the phase difference $\phi_1-\phi_2$ which is proportional to the applied magnetic flux piercing the loop area (red area in c). The latter enters via the sum of phases $\phi_1+\phi_2$, when choosing a gauge such that the time-derivative is proportional to the induced electric field (via the Faraday effect). For the single junction circuit (b) only the EMF mode exists. (d) Equivalent circuit diagram for the dc SQUID in (c).}
    \label{fig1_main}
\end{figure*}

Since spins, or more generally, magnetic dipoles exhibit a magnetic field with strong spatial dependence including a singular origin, their precise modelling remains a major challenge. The most popular approach~\cite{Bialczak_2007,Lanting_2014,LaForest_2015,Aquino_2022} attempts to circumvent this problem by invoking the principle of reciprocity: instead of considering the dipole magnetic field felt by Cooper pairs moving inside the superconducting circuit, one considers the magnetic field felt by the dipole induced by loop supercurrents. However, as we argue in this work, this treatment is \textit{incomplete} due to Faraday's law of induction. Time-varying magnetic fields $\mathbf{B}(t)$ inescapably introduce an associated \textit{electric field} via Faraday's law, $\nabla\times\mathbf{E}=-\dot{\mathbf{B}}$. These two fields cannot be treated separately. Indeed, their connection is engrained in the underlying gauge structure of quantum Hamiltonians~\cite{You_2019,riwar_2022}: a given vector potential $\mathbf{A}$ encoding the magnetic field ${\mathbf{B}=\nabla\times\mathbf{A}}$ can always be chosen in a gauge such that it automatically represents also the electromotive field via $\mathbf{E}_\mathbf{\dot{B}}=-\dot{\mathbf{A}}$. Other gauge choices are by all means possible by applying a unitary transformation $U$, but this transformation yields the extra quantum geometric term $-iU\dot{U}^\dagger$ in the Schrödinger equation~\cite{Messiah1961}, preserving the electro-motive force (EMF).

While this quantum geometric connection to the Faraday effect has recently been noticed~\cite{You_2019,riwar_2022}, many important aspects have not been dealt with, such as 2D thin film geometries and finite size effects of the capacitor wings. Moreover, there is one crucial feature which has to the best of our knowledge not been analysed at all. Time-varying fluxes couple in general even to circuits \textit{without flux loops} via the induced electromotive field, see also Fig.~\ref{fig1_main} -- an effect which is phenomenologically speaking indistinguishable from regular dielectric loss. While the absence of flux-tunability indeed leaves the qubit frequency constant, the EMF provides a nonzero matrix element coupling ground and excited qubit states, and can thus contribute to the quality factor via a regular qubit relaxation process.

The present work is dedicated to quantifying this simple observation, and setting up a general formalism to describe the coupling of magnetic fields and corresponding electromotive fields to superconducting thin films, enabling both analytic and numerical model calculations for typical transmon device geometries. While this formalism is based on Ref.~\cite{riwar_2022}, we significantly extend on aspects related to thin films, and connect charge conservation of the surface charges screening the electromotive field to Gauss' law for the magnetic field. We furthermore compare our framework to the reciprocal treatment in the existing literature~\cite{Bialczak_2007,Lanting_2014,LaForest_2015,Aquino_2022}, and find that the latter is of limited use. It is not so much a matter of validity of the reciprocity principle, as it is a failure to express the EMF mode in terms of a well-defined equivalent supercurrent.

We then apply this framework to magnetic dipoles, which can describe substrate or surface spins, as well as the far field of flux control lines. To summarise, we find the following. Assuming that surface spins (coupling locally to the junction bridges) are indeed at the microscopic origin of $1/f$ flux noise, there is a fundamental upper limit to the qubit quality factor somewhere between $10^8\sim 10^{10}$. Depending on the spin-spin correlation length, and how this correlation length evolves from the sub-Hz to GHz regime, the qubit quality factor limit may suffer a further reduction by at least an order of magnitude, due to a long-range contribution from the transmon capacitor wings. We then consider far-field contributions due to current-carrying flux lines, where we predict two mechanisms contributing to a relatively long range coupling. On the one hand, the EMF decays only with $1/r^2$ ($r$ is the distance between qubit and source) whereas the magnetic field decays with $1/r^3$. On the other hand, the coupling is amplified by a lensing of the electromotive field due to the transmon capacitor wings. This lensing effect is distinct from usual Meissner flux focussing and instead pertains to the interplay between longitudinal and transversal electric fields at the interface between superconducting material and free space. As a matter of fact, based on a type of holographic property for the electromotive force at the superconducting surfaces, we find that for thin films, Meissner screening plays a minor role independent of the Pearl length.

Overall, our work provides a highly general framework to describe the coupling between magnetic field sources and superconducting thin films. It highlights flux noise as a detrimental ingredient \textit{even for fixed-frequency} superconducting qubits, indicating that with record quality factors of  $10^7\sim 10^8$~\cite{Bland_2025,Dane_2025} the field might soon close in on a fundamental limit relevant for generic charge qubits. 

This work is organized as follows. Section~\ref{sec_charge_qubit_revisited} revisits the standard charge qubit in the context of the Faraday effect. We then outline the here presented irrotational formalism to account for the Faraday effect in superconducting thin films in Secs.~\ref{sec_Faraday_effect_thin_films} and~\ref{sec_surface_charges_conformal}. Some basics of magnetic dipole ensembles are summarised in Sec.~\ref{sec_magnetic_dipoles}. After a short comparison between the here proposed formalism and the reciprocal treatment known from the existing literature in Sec.~\ref{sec_limitations_reciprocity}, we present two applications. In Sec.~\ref{sec_brigdes_vs_wings} we consider surface spin ensembles, and analyse their impact on the quality factor limit. Finally, in Sec.~\ref{sec_safety_distance} we consider the coupling to flux lines and introduce a minimal safety distance for the qubit to remain unaffected by flux control.

\section{Charge qubit revisited}\label{sec_charge_qubit_revisited}

Consider a standard charge qubit either with a single junction, Fig.~\ref{fig1_main}(b), or with a dc SQUID loop, Fig.~\ref{fig1_main}(c). When reduced to a lumped-element circuit, see Fig.~\ref{fig1_main}(d), the device dynamics is captured by a Hamiltonian containing a single node with the canonically conjugate charge-phase pair $[\widehat{\phi},\widehat{N}]=i$
\begin{align}\label{eq_Hamiltonian_SQUID}
H&=E_{C}\left(\widehat{N}+N_g\right)^{2}- E_{J1}\mathrm{cos}\left(\widehat{\phi}+\phi_{1}\right) \nonumber \\ 
&-E_{J2}\mathrm{cos}\left(\widehat{\phi}+\phi_{2}\right), 
\end{align}
with the charging energy $E_C$, the offset gate charge $N_g$, and the Josephson energies $E_{J1}$ and $E_{J2}$ of the two junctions comprising the loop. The single junction device follows by simply setting $E_{J2}=0$. We assume the device to be operated in the transmon regime, where the large capacitor wings significantly reduce the charging energy $E_{J1},E_{J2}\gg E_C$. Depending on whether there is a single junction or two junctions, this circuit realises a fixed-frequency or flux-tunable qubit, both staple elements in various contemporary quantum hardware setups. The central subject of interest in this work is the interaction of this circuit with any externally applied flux. Formally, this coupling is captured with the two phases $\phi_1$ and $\phi_2$ entering the Josephson junctions (equivalent to Peierls phases). The main challenge is the correct (and gauge invariant) computation of $\phi_{1,2}$.

The importance of gauge aspects can be appreciated as follows. If the applied magnetic field is time-independent, the only gauge invariant quantity is the flux enclosed by the loop,
\begin{equation}\label{eq_loop_flux}
\delta\phi=\phi_1-\phi_2\ .
\end{equation}
As a consequence, the single-junction device does not couple to static magnetic fields.
In recent years, there has emerged a heightened awareness of the fact that the generalisation to \textit{time-varying} fluxes (relevant for deterministic driving, as well as fluctuating noise) is not trivial~\cite{You_2019,riwar_2022}, because of the extra electric field induced by the magnetic field via Faraday's law of induction, $\nabla\times\mathbf{E}=-\dot{\mathbf{B}}$. Translated to the language of the above SQUID circuit, the issue is that two different sets of phase drops $\phi_1,\phi_2$ and $\phi_1^\prime,\phi_2^\prime$ (both satisfying $\delta\phi=\phi_1-\phi_2=\phi_1^\prime-\phi_2^\prime$) are connected by a time dependent unitary transformation $U(t)$ which yields the extra term $-iU\dot{U}^\dagger$ in the Schr\"odinger equation~\cite{Messiah1961}, such that the loop constraint of Eq.~\eqref{eq_loop_flux} alone does not suffice to correctly account for all electromagnetic fields.

Let us make this observation explicit by simplifying the Hamiltonian for the special case of a symmetric SQUID, $E_{J1}=E_{J2}\equiv E_J$. We find via standard trigonometric identities, and the time-dependent unitary transformation $U=e^{-i(\phi_1+\phi_2)\widehat{N}/2}$ the Hamiltonian $\widetilde{H}=UHU^\dagger -iU\dot{U}^\dagger$, with (see Appendix~\ref{app_SQUID_transformation})
\begin{equation} \label{eq_Hamiltonian_SQUID_trafo}
\widetilde{H}=E_C(\widehat{N}+N_g)^2+\frac{\dot{\phi}_{1}+\dot{\phi}_{2}}{2}\widehat{N}-E_{J,\text{loop}}\cos\left(\widehat{\phi}\right),
\end{equation}
with the effective loop Josephson energy $E_{J,\text{loop}}=2E_{J}\cos\left(\frac{\phi_{1}-\phi_{2}}{2}\right)$. Overall, we can decompose the two phases into a flux mode $\delta \phi= \phi_1-\phi_2$ describing the regular Aharonov-Bohm effect, and a so-called EMF mode, $ \phi_1+\phi_2$, whose time-derivative describes the electric field induced by the Faraday effect. The former renders the qubit eigenfrequency $\omega_0\approx \sqrt{2E_{J,\text{loop}}E_C}$ (transmon limit) flux-tunable, whereas the latter effectively provides a dynamic contribution to the offset charge $N_g$.

In order to describe losses let us consider an ensemble of fluctuating flux sources.
In that scenario, the phase drops $\phi_{1,2}$ acquire an operator component,
\begin{align}\label{eq_phi_to_phi_hat}
    \phi_{1,2}\rightarrow \phi_{1,2}+\widehat{\phi}_{1,2}\ ,
\end{align}
where $\phi_{1,2}$ account for a classical, deterministic control flux, whereas the operators $\widehat{\phi}_{1,2}$ account for the ensemble responsible for flux noise. The dynamics of these fluctuations are formally captured by including an environment Hamiltonian, $H\rightarrow H+H_\text{env}$.

The relaxation and dephasing rates can be computed by Fermi's Golden rule, and assume the form (see Appendix~\ref{app_relaxation_and_dephasing})
\begin{align}\label{eq_Gamma_1}
   \Gamma_{1}= & \sqrt{\frac{E_{J}}{E_{C}}}\frac{\omega_{0}^{2}}{4}\left[S_{11}(\omega_0)+2S_{12}(\omega_0)+S_{22}(\omega_0)\right]\ , \\ \label{eq_Gamma_varphi}
   \Gamma_{\varphi}= & \frac{\left(\partial_{\delta \phi}\omega_0\right)^{2}}{2}\left[S_{11}(0)-2S_{12}(0)+S_{22}(0)\right], 
\end{align}
with the symmetrised power spectral densities (PSD),
\begin{equation}
S_{\nu,\nu^\prime}\left(\omega\right)=\frac{1}{2}\int_{-\infty}^{\infty}\mathrm{d}te^{i\omega t}\left\langle \left\{ \widehat{\phi}_\nu\left(t\right),\widehat{\phi}_{\nu^\prime}\left(0\right)\right\} \right\rangle\ , \label{eq:PSD} 
\end{equation}
where the phase operators are in the interaction picture $\widehat{\phi}_\nu(t)=e^{iH_\text{env}t}\widehat{\phi}_\nu e^{-iH_\text{env}t}$, and we assumed the phase-phase correlator to be symmetric upon exchanging $\nu \leftrightarrow \nu^\prime$.

As a side note, we point out that in Fermi's golden rule, the dephasing $\Gamma_\varphi$ is nominally measured at zero frequency (hence putting the frequency argument inside the correlators to $0$). In practice, however, dephasing measurements (such as, e.g., free induction decay or Hahn echo) involve pulses within a finite measurement time, whose details are encoded by means of a frequency filter function~\cite{Cywinski_2008}. For free decay as an example, one can effectively replace the exponent for the loss of coherences, $e^{-\Gamma_\varphi t}$, as
\begin{equation}
    \Gamma_\varphi t \rightarrow \int_0^\infty \frac{d\omega}{\pi}\frac{2\sin^2(\omega t/2)}{\omega^2}\Gamma_\varphi(\omega)\ ,
\end{equation}
where $\Gamma_\varphi(\omega)$ is defined as in Eq.~\eqref{eq_Gamma_varphi}, except that the correlators are taken at finite frequency, $S_{\nu\nu^\prime}(0)\rightarrow S_{\nu\nu^\prime}(\omega)$.

The pure dephasing term $\Gamma_\varphi$ is due to fluctuations in the regular flux mode $\phi_1-\phi_2$, which in turn lead to fluctuations in the qubit frequency. The EMF mode, on the other hand, yields (via coupling to the charge operator $\widehat{N}$) a finite transition matrix element between the qubit states, thus leading to a relaxation process with nonzero $\Gamma_1$. Note that while $S_{\nu,\nu^\prime}(\omega)$ represents the phase-phase correlators, for $\Gamma_1$ one nominally needs to compute the correlators of the time-derivatives of the phases, $\partial_t \widehat{\phi}_\nu$. This can be included very easily in Fourier space by multiplication with the frequency, $S_{\nu,\nu^\prime}(\omega)\rightarrow \omega^2 S_{\nu,\nu^\prime}(\omega)$, which is why in $\Gamma_1$ of Eq.~\eqref{eq_Gamma_1}, the extra prefactor $\omega_0^2$ appears.

The existence and relevance of the latter (relaxation) process, and its connection to gauge transformations, was first elaborated by Ref.~\cite{You_2019} in a pure lumped-element language. In this language, the authors assigned to each junction a self-capacitance, $C_1$ and $C_2$, and found that $\phi_1=C_2\delta\phi/C_\text{tot}$ and $\phi_2=-C_1\delta\phi/C_\text{tot}$, where $\delta\phi$ is proportional to the flux enclosed by the SQUID loop area. According to this model, there still only can be a nonzero coupling between flux sources and the qubit, if there is a finite loop. For a single junction, $\delta\phi$ must be zero, such that $\phi_1=\phi_2=0$. Hence, in a simplified lumped-element treatment, there cannot be a flux-noise induced relaxation process without flux-tunability (unless one included the Fraunhofer effect, which is weak for the here considered small junctions). However, Ref.~\cite{riwar_2022} subsequently pointed out that in general, the lumped-elements perspective is not enough, and accurate modelling requires knowledge of the detailed device geometry and flux distribution in continuous position space -- especially in the case of the transmon, where the dominant part of the circuit's capacitance is not provided by the individual junction self-capacitances, but by large capacitor wings. These wings have been shown to be the decisive component regarding the screening of surface charges due to $\mathbf{E}_{\dot{B}}$, which in turn dominate the coupling strength between the circuit and a given flux source~\cite{riwar_2022}.

In Ref.~\cite{riwar_2022} it was found in particular that $\phi_1$ and $\phi_2$ are for generic device geometries not simply related to the loop flux $\delta\phi=\phi_1-\phi_2$ by a prefactor (as opposed to the lumped-element treatment of~\cite{You_2019}), and instead $\phi_1$ and $\phi_2$ can be independent functions of time. While not explicitly stated nor explored in Ref.~\cite{riwar_2022}, this finding implies that there is no fundamental reason to exclude the possibility of fluxes coupling to (non flux-tunable) single junction circuits. As a matter of fact, all that is required is that the electromotive field due to the Faraday effect be nonzero at the junction. As we will show in what follows, this is actually the default scenario in general.

In fact, we can give a rough but concrete first estimate already at this stage. Under the assumption that the flux correlations are local, such that $S_{11}(\omega),S_{22}(\omega)\gg S_{12}(\omega)$, we take the expression for the relaxation rate from Eq.~\eqref{eq_Gamma_1}, the universal $1/f$ flux noise hypothesis $S_{11}(\omega)\approx 4\pi^2 A^2/(\omega \Phi_0^2)$ (the $4\pi^2/\Phi_0^2$ prefactor comes from the relationship between superconducting phase and flux, $\phi=2\pi \Phi/\Phi_0$), and express the quality factor $Q$ accordingly as the ratio between qubit frequency and relaxation rate
\begin{equation}\label{eq_Q_limit}
    Q=\frac{\omega_0}{\Gamma_1}\approx \sqrt{E_C}{E_J}\frac{\Phi_0^2}{\pi^2 A^2}\ .
\end{equation}
With the transmon regime $\sqrt{E_C/E_J}\sim 10^{-1}$, the prefactor $1/\pi^2\sim 10^{-1}$, and the flux noise amplitude $A$ ranging between $10^{-5}\sim 10^{-6}\Phi_0$, we get a \textit{fundamental (universal) upper limit} of the quality factor between $Q\sim 10^8\sim 10^{10}$ due to flux noise. Again, we stress that this estimate is valid for both fixed frequency (single junction) and flux-tunable transmons. For a single junction, only $S_{11}$ is nonzero. Here, flux noise actually provides only relaxation but no dephasing, since Eq.~\eqref{eq_Gamma_varphi} yields zero due to $\partial_{\delta\phi}\omega_0 =0$.


This estimate seems particularly relevant considering the very recent technological leap in transmon qubit quality factors, reaching the regimes $10^7\sim 10^8$~\cite{Bland_2025,Dane_2025}. If correct, it implies that further improvements on transmon quality factors might soon require an engineering effort to reduce flux noise to amplitudes below its currently known universal value. In what follows, we develop a detailed framework allowing for quantitative modelling of the coupling between circuits and magnetic fields including the Faraday effect. This will allow us to justify the above estimate in more detail (by considering the explicit coupling to spin ensembles close to the device). In fact, under certain parameter regimes, there may be scenarios, where the quality factor upper limit receives a further reduction. Furthermore, we will be able to consider far-field contributions due to current-carrying flux lines on the same footing, and introduce a so-called "safety radius" (i.e., a critical distance) between Josephson junctions and flux lines, below which the lines have a further detrimental effect on the quality factor -- again valid likewise for both fixed-frequency and tunable transmons.

\section{Faraday effect in thin films} \label{sec_Faraday_effect_thin_films}

As we have just established, it is important to account for both the magnetic and the corresponding induced electric field for an accurate description of qubit losses. Given a device geometry and magnetic field distribution as input, Ref.~\cite{riwar_2022} provided a concrete recipe to compute the phase drops at the junctions $\phi_\nu$, which we here review and extend (specifically geared towards thin film devices).

\begin{figure*}
    \centering
    \includegraphics[width=1.0\linewidth]{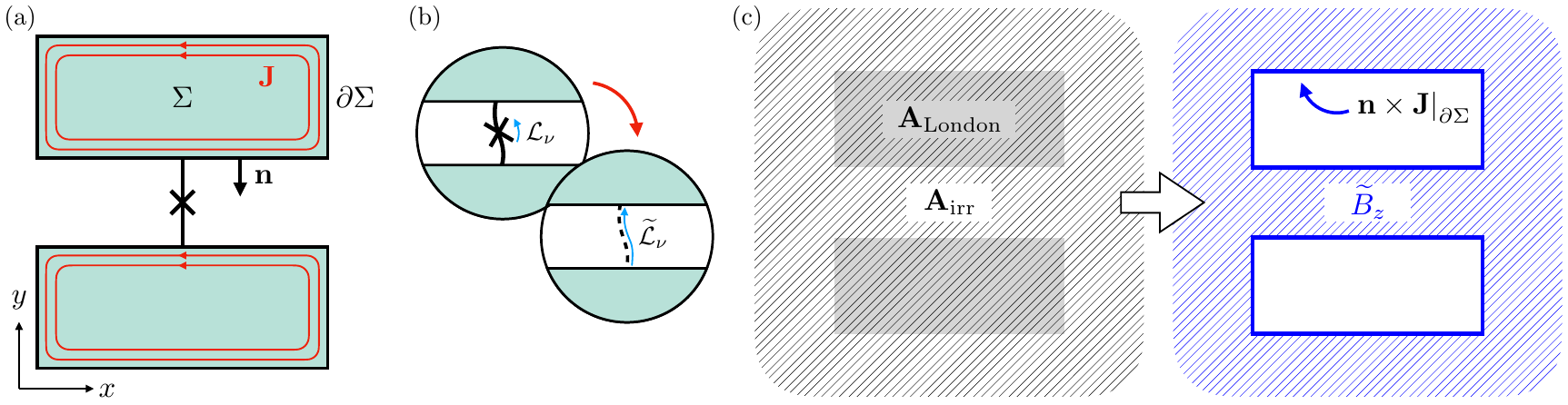}
    \caption{Aspects relevant for the formulation of the irrotational gauge for the vector potential $\mathbf{A}$, which allows to compute the phase drops $\phi_\nu$ in Fig.~\ref{fig1_main}. Top view (plane in $x$ and $y$) of a transmon device (a). The large capacitor wings form a thin film bulk $\Sigma$, with boundary $\partial \Sigma$ and the unit vector normal to the boundary $\mathbf{n}$. Due to Meissner (Pearl) screening in 2D, a superconducting sheet current $\mathbf{J}$ flows within the thin film. The filamentary bridge forming the Josephson junction can be neglected in the calculation of all relevant fields due to the large capacitor wings (b), see also Ref.~\cite{riwar_2022}. The computation of the irrotational vector potential $\mathbf{A}_\text{irr}$ (in free space), as well as the 2D Meissner screening problem, can be simplified by introducing the effective magnetic field $\widetilde{B}_z$ (c). This effective field is defined to be zero inside the bulk superconductors ($\Sigma$) and equal to $B_z$ in free space (blue hatched area), plus a delta-like contribution due to the screening supercurrent $\sim \mathbf{n}\times\mathbf{J}$ on the rim $\partial\Sigma$ (blue solid lines). The delta-like edge contribution has a holographic quality, as it contains the information of the total flux penetrating the bulk (captured by the vector potential in the London gauge $\mathbf{A}_\text{London}$).}
    \label{fig2_btilde}
\end{figure*}

Consider a transmon layout as in Fig.~\ref{fig2_btilde}(a) with a thin bridge containing the Josephson junction, and large capacitor wings (made of large thin film layers, which we refer to as the bulk volume $\Sigma$ in what follows). The main idea consists of finding a representation (gauge fixing) for the vector potential such that it captures both the time-varying magnetic field and the corresponding induced electric field, $\mathbf{B}=\nabla\times\mathbf{A}$ and $\mathbf{E}_{\dot{B}}=-\dot{\mathbf{A}}$. With the magnetic field $\mathbf{B}$ we mean the full field including the Meissner effect (more on that later). Within the superconductor (volume $\Sigma$), the textbook London gauge ($\mathbf{A}_\text{London}$) fulfills the above criteria. In free space, this gauge is called the `irrotational gauge' ($\mathbf{A}_\text{irr}$), adopting the terminology of Ref.~\cite{You_2019,riwar_2022}. In this gauge, the phase drop $\phi_\nu$ at junction $\nu=1,2$ appearing inside the Hamiltonian of Eq.~\eqref{eq_Hamiltonian_SQUID} is simply computed via the line integral
\begin{align}
    \phi_{\nu}=\frac{2\pi}{\Phi_{0}}\int_{\mathcal{L}_{\nu}}\mathrm{d}\mathbf{l}\cdot\mathbf{A}_\text{irr}, \label{eq:irr_gauge_RD}
\end{align}
where $\mathcal{L}_\nu$ is the path covered by Cooper pairs as they travel across the $\nu$th junction.

Importantly, in the presence of large capacitor wings, the junction self-capacitances can be neglected. This significantly simplifies the problem. One simply computes $\mathbf{A}_\text{irr}$ explicitly for the large superconducting structures without the actual junction bridges (the Niemeyer-Dolan filaments). The information of the bridge path is then included in Eq.~\eqref{eq:irr_gauge_RD} by replacing the line integral $\mathcal{L}_\nu$ (going over the actual, physical junction) by the line integral $\widetilde{\mathcal{L}}_\nu$ going over the entire bridge, see also Fig.~\ref{fig2_btilde}(b). Note importantly, that this step requires the simplifying assumption that the junction bridges (filaments) can be assumed to be one-dimensional. When considering magnetic dipoles situated close to the bridge (see further below) we will deviate from this assumption and include a finite bridge width (i.e., the bridge is still 2D, but has a finite width $b$ within the plane), in order to regularise divergences.

The irrotational gauge furthermore has to satisfy three conditions (i-iii) in order to be uniquely determined. In the free space (outside of the superconducting bodies), it fulfills the Coulomb gauge $\nabla\cdot \mathbf{A}_\text{irr}=0$ (i). At the superconducting surfaces ($\partial\Sigma$), the transversal part (parallel to the surface, i.e., perpendicular to the normal vector $\mathbf{n}$) has to be continuous with the London gauge in the superconducting interior (ii). The longitudinal part (parallel to $\mathbf{n}$), on the other hand, can be (and in general is) discontinuous. This discontinuity gives rise to the surface charges which screen the induced electric field $\mathbf{E}_{\dot{B}}$. For each connected superconducting body $\Sigma$, this surface charge needs to integrate to zero (iii). Via the relationship $\mathbf{E}=-\dot{\mathbf{A}}_\text{irr}$, (iii) imposes that the surface integral of the part of $\mathbf{A}_\text{irr}$ normal to the surface $\partial\Sigma$ equally integrates to zero.

We here add the following clarifications to the above recipe, relevant specifically for thin film layers with thickness $d$ (which have not been explicitly analysed in Ref.~\cite{riwar_2022}). Since the supercurrent is constrained to move in the 2D plane ($x$ and $y$), the thin film superconductor dominantly couples to the $z$-component of the magnetic field, $B_z$, pointing out of the plane. Integrating over the thin layer $d$, it is useful to replace the usual supercurrent density $\mathbf{j}$ (responsible for Meissner screening) by a 2D sheet current  $\mathbf{J}=d\mathbf{j}$ (Amp\`{e}re per length). With this redefinition, we get that within the thin film, the London gauge has to satisfy
\begin{equation}\label{eq_A_london}
    \mathbf{A}_\text{London}=-\frac{\mu_0 \Lambda}{2}\mathbf{J}\ ,
\end{equation}
where the Pearl length $\Lambda=2\lambda^2/d$ (relevant for Meissner screening in 2D) naturally appears. Recent experiments reported on $\Lambda\sim 2\mu\text{m}$ for $\sim 25\text{nm}$ aluminium films~\cite{LopezNunez_2025}. In comparison, capacitor wings in typical transmon designs (see, e.g., micrographs in Refs.~\cite{Riwar_2016b} and~\cite{kim2025}) commonly have a shorter dimension of $\sim 10 \mu\text{m}$ and a longer dimension of at least $\sim 100 \mu\text{m}$ up to $\sim 1 \text{mm}$. The 2D curl $\nabla\times \mathbf{A}=\partial_y A_x-\partial_x A_y$ for the above London gauge automatically returns the corresponding field $B_z$. Within $\Sigma$, the sheet current furthermore has to satisfy $\nabla\cdot \mathbf{A}_\text{London}=0$ (just like the irrotational component), and at the boundary $\left.\mathbf{n}\cdot \mathbf{A}_\text{London}\right\vert_{\partial\Sigma}=0$, and is thus fully determined by the above condition for singly connected volumes (genus 0).

Note that the London gauge within the bulk only enters effectively as a boundary condition for the free space solution $\mathbf{A}_\text{irr}$. This allows us to include condition (ii) by means of a convenient trick, which we use for explicit computations further below. Instead of considering the field $B_z$, we introduce the field
\begin{equation}\label{eq_btilde}
    \widetilde{B}_z=B_z+\frac{\mu_0 \Lambda}{2}\delta_{\partial\Sigma}(\mathbf{r})\mathbf{n}\times\mathbf{J}\ ,
\end{equation}
where $\delta_{\partial\Sigma}(\mathbf{r})$ is a Dirac delta distribution nonzero for positions $\mathbf{r}$ along the edge $\partial\Sigma$. To get the field relevant to the calculation of the irrotational gauge, we take $\widetilde{B}_z$ within free space ($\mathbf{r}\in\mathbb{R}^2\setminus \Sigma$, i.e., all of $\mathbb{R}^2$ except for $\Sigma$), and zero field within the bulk ($\mathbf{r}\in\Sigma$), see Fig.~\ref{fig2_btilde}(c). Since the delta distribution is placed exactly on the rim, it is useful to introduce the notion of the areas $\Sigma^{\pm}$, which corresponds to the thin film area $\Sigma$ either including the rim and thus the delta distribution ($\Sigma^+$) or excluding it ($\Sigma^-$). Consequently, for any function $f(\mathbf{r})$,
\begin{equation}
\begin{split}
    \iint d^2r \left(\widetilde{B}_z-B_z\right)f= \iint_{\Sigma^+} d^2r \left(\widetilde{B}_z-B_z\right)f\\=\frac{\mu_0\Lambda}{2}\oint dr \left(\mathbf{n}\times \mathbf{J}\right)f\ ,
\end{split}
\end{equation}
whereas $\iint_{\Sigma^-} d^2r (\widetilde{B}_z-B_z)f=0$. The construction of $\widetilde{B}_z$ (when excluding $\Sigma^-$) can be connected to a type of `holographic' principle. In particular, note that via Eq.~\eqref{eq_A_london} and Stokes' theorem, one can show for any integration area $\mathcal{A}$ that includes the entirety of a superconducting volume $\Sigma^+$ ($\Sigma^+ \subset \mathcal{A}$), that
\begin{equation}\label{eq_holographic}
    \iint_{\mathcal{A}\setminus\Sigma^-} d^2r \widetilde{B}_z=\iint_\mathcal{A} d^2r B_z\ .
\end{equation}
We refer to this feature as holographic in the sense that the entire information of the flux penetrating the superconducting volume $\Sigma$ is contained within the boundary contribution given in Eq.~\eqref{eq_btilde}. This feature will be used below. Note that while we usually denote with $\Sigma$ the entirety of the superconducting bulk (i.e., for the transmon both upper and lower capacitor plate), Eq.~\eqref{eq_holographic} also works when taking only a single connected superconducting bulk.

It is further interesting to note that the exact same construction $\widetilde{B}_z$ also appears within the Meissner problem itself. When integrating out the third dimension ($z$) due to the thin film setup, it is no longer possible to write the London equations in terms of pure differential equations. Instead, the screened magnetic field has to be computed with an integral equation, which reads
\begin{equation}\label{eq_Meissner_2D}
    B_{z}=B_{z}^{(0)}-\frac{1}{2\pi\Lambda}\iint_{\Sigma^+}d^{2}r^{\prime}\frac{\widetilde{B}_{z}\left(\mathbf{r}^{\prime}\right)}{\left|\mathbf{r}-\mathbf{r}^{\prime}\right|}\ ,
\end{equation}
where the bare field of the flux source is denoted as $B_{z}^{(0)}$. Naturally, since screening is due to a nonzero Cooper pair density, the integral is limited to the superconducting bulks $\Sigma$. Importantly, in this integral equation, there emerges the same boundary term $\sim \mathbf{n}\times\mathbf{J}$ due to the interplay between the second London equation and Amp\`{e}re's law in the integral form (valid for $\nabla\cdot\mathbf{j}=0$),
\begin{equation}\label{eq_ampere}
    \mathbf{B}=-\frac{\mu_{0}}{4\pi}\iiint d^{3}r^{\prime}\frac{\left(\mathbf{r}-\mathbf{r}^{\prime}\right)\times\mathbf{j}\left(\mathbf{r}^{\prime}\right)}{\left|\mathbf{r}-\mathbf{r}^{\prime}\right|^{3}}\ ,
\end{equation}
via reduction to 2D ($\iiint d^{3}r^{\prime}\rightarrow \iint d^{2}r^{\prime}$ and $\mathbf{j}\rightarrow\mathbf{J}$) and a partial integration step. We here include this boundary term by means of $\widetilde{B}_z$, Eq.~\eqref{eq_btilde}, and setting the integration volume to $\Sigma\rightarrow \Sigma^+$. At any rate, this boundary term guarantees that the 2D equivalent of the Gauss law, $\iint d^2 r B_z=0$, is preserved for any solution $B_z$ of Eq.~\eqref{eq_Meissner_2D}, as long as the input (bare) field satisfies it, $\iint d^2 r B_{z}^{(0)}=0$. While in 3D, the Meissner effect yields the typical exponential suppression of the magnetic field on the scale of the London penetration depth $\lambda$, the 2D Meissner problem is well-known to yield a weaker algebraic suppression. Nonetheless, devices larger than $\Lambda$ can provide significant distortion of the magnetic field. It is therefore interesting to note that for the computation of $\phi_\nu$, we will find further below that 2D Meissner screening has in general a minor effect even in the case of small $\Lambda$, due to the aforementioned holographic feature represented in $\widetilde{B}_z$.

\begin{figure}
    \centering
    \includegraphics[width=\linewidth]{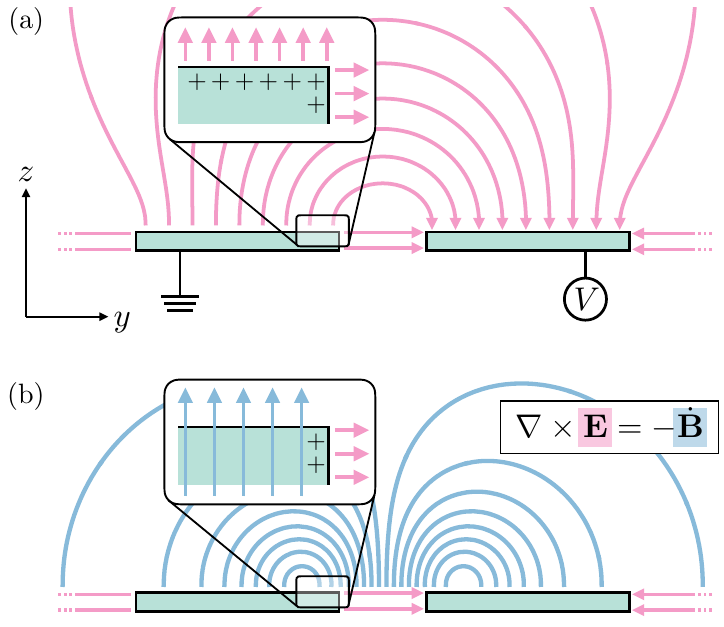}
    \caption{Justification for neglecting the top and bottom surfaces of the thin film for the Faraday problem. For regular electrostatics in voltage-biased co-planar capacitors, top and bottom surface charges are dominant (a). In contrast, for a magnetic field with dominant $z$-component (perpendicular to the chip plane), the corresponding induced electric field is in-plane, leading to dominant surface charges at the 1D rim (b).}
    \label{fig3_surface_charges}
\end{figure}

For numerical evaluations of Eq.~\eqref{eq_Meissner_2D}, we resort to an iterative approach. We first start with the bare field $B_{z}^{(0)}$ (given for concrete cases below, such as a magnetic dipole), compute the corresponding sheet current $\mathbf{J}^{(0)}$ via the London gauge, Eq.~\eqref{eq_A_london}, construct $\widetilde{B}_{z}^{(0)}$ as in Eq.~\eqref{eq_btilde}, and then insert it into the Pearl kernel $\sim 1/\Lambda$. This yields a new output field $B_z^{(1)}$, which undergoes the same steps, and is reinserted, until the output field converges. Conceptually, this corresponds to solving the iterative equation
\begin{equation}\label{eq_Meissner_iteration}
    B_{z}^{(n)}=B_{z}^{(n-1)}-\frac{1}{2\pi\Lambda}\iint_{\Sigma^+}d^{2}r^{\prime}\frac{\widetilde{B}_{z}^{(n-1)}\left(\mathbf{r}^{\prime}\right)}{\left|\mathbf{r}-\mathbf{r}^{\prime}\right|}\ .
\end{equation}
Depending on the specifics of the input field, a direct implementation of this equation might run into convergence problems. This problem can be fixed by introducing a mixing procedure, where the updated iteration is replaced by a weighted average with the previous iteration, $B_{z}^{(n)}\rightarrow m B_{z}^{(n)}+(1-m) B_{z}^{(n-1)}$, before reinserting it into the right-hand side. The mixing parameter $m$ has to be chosen empirically upon tracking the convergence behaviour. In order to render the evaluation of the convolution with the Pearl kernel efficient, we go to Fourier space via a fast Fourier transform (FFT, discretizing both real space and the corresponding Fourier space). In Fourier space, the Pearl kernel reads $1/(\Lambda |\mathbf{k}|)$ where the divergent zero mode $\mathbf{k}=0$ is omitted. The same FFT is also used in the computation of $\mathbf{J}$. In order to guarantee $\nabla\cdot\mathbf{J}=0$ we resort to a standard stream function approach.

Note that the above reduction to 2D only takes into account screening surface charges on the 1D rim of the boundary $\partial \Sigma$. In principle, there could also be surface charges on the top and bottom surfaces of the thin films, when changing the view again to 3D. In fact, such top and bottom surface charges give rise to the main relevant capacitive coupling for regular electrostatics in co-planar capacitors with a voltage bias, see Fig.~\ref{fig3_surface_charges}(a). However, for the screening of the electromotive force (due to $\dot{\mathbf{B}}\neq 0$), the position of surface charges strongly depends on the orientation of the applied magnetic field. If the field has a dominant $z$-component (normal to the thin film), then indeed the orientation of the $\mathbf{E}$-field induced by Faraday's law is mostly in-plane, and surface charges are located at the 1D rim, see Fig.~\ref{fig3_surface_charges}(b). For magnetic dipole moments (due to surface spins on the superconductor or the substrate, or due to flux lines) the dominant coupling indeed comes from the $z$ component of the dipole, justifying the above assumption, as we discuss in more detail further below.


\section{Surface charges and conformal maps}\label{sec_surface_charges_conformal}

Overall, the sequence to get to the correct coupling $\phi_\nu$ goes as follows. One first identifies a given external magnetic field source $b_z$, then solves for 2D Meissner screening via Eq.~\eqref{eq_btilde} to get $B_\text{z}$ and then uses the recipe detailed above to compute $\mathbf{A}_\text{irr}$, which is inserted into Eq.~\eqref{eq:irr_gauge_RD} with $\mathcal{L}_\nu$ representing the path of the junction filament.

\begin{figure*}
    \centering
    \includegraphics[width=0.8\linewidth]{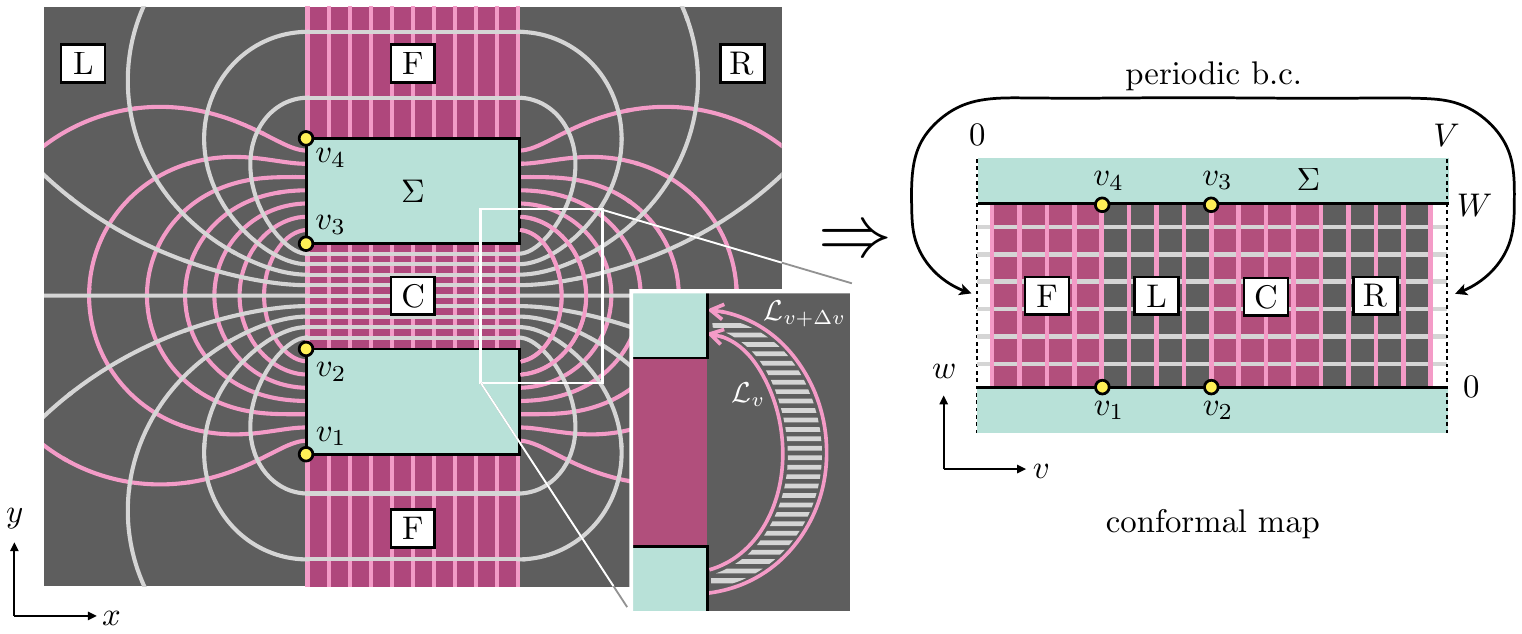}
    \caption{Exploiting conformal maps to compute the vector potential in the irrotational gauge $\mathbf{A}_\text{irr}$ and ultimately the phases $\phi_\nu$. We illustrate the principle by means of the device geometry in Fig.~\ref{fig1_main}, representing a transmon with rectangular capacitor wings. The real space (left panel) excluding the superconducting volumes $\Sigma$ is mapped onto a rectangle with periodic boundaries (cylinder manifold, right panel). For rectangular capacitor wings, the regions C, L, R, and F are mapped piece-wise, and stitched together via continuity constraints. A simplified computation for $\phi_\nu$ is possible for the case when bridge paths are straight lines in the conformal space, i.e., with coordinate $v$ constant and $w$ going from $0$ to $W$, by computing the flux through slices with neighbouring values $v$ and $v+\Delta v$ (inset).}
    \label{fig4_conformal}
\end{figure*}

A final challenge in the calculation of $\mathbf{A}_\text{irr}$ concerns the correct implementation of condition (iii), i.e., the requirement that surface charges along the boundary $\partial\Sigma$ integrate to zero. While Ref.~\cite{riwar_2022} already provided closed integral expressions for $\phi_\nu$ satisfying condition (iii) for some example geometries, those geometries involved simplified parallel capacitor plates where fringe effects and surface charges on the far side of the plates were neglected. We here strive to give a realistic account for thin film co-planar setups, including surface charges everywhere on the device, an ingredient that will be relevant further below.

This ingredient can be conveniently taken into account (for capacitor wings with genus 0) by performing a conformal map from the entire free space excluding $\Sigma^-$ ($\mathbb{R}^2\setminus\Sigma^-$) to a cylinder manifold (rectangular space with periodic boundaries), see Fig.~\ref{fig4_conformal}. This general problem can be solved with ease in the conformal space by means of a discrete Fourier approach, exploiting the periodic boundaries in one direction ($w$), and the Dirichlet boundary condition (zero $\widetilde{B}_z$ inside the volume $\Sigma^-$) in the other direction ($v$). Let us denote the conformal map as the map from real space $\mathbf{r}=(x,y)$ to the conformal space $\mathbf{u}=(v,w)$ (and the boundaries of the rectangular strip in conformal space shall be at $v=0,V$ and $w=0,W$). In this conformal space, the general solution for the irrotational vector potential reads formally
\begin{equation}\label{eq_Airr_conformal}
\begin{split}
    (\mathbf{A}_\text{irr})_{v,w}=\int_0^V dv^\prime \int_0^W dw^\prime K_{v,w}(v-v^\prime;w,w^\prime)\\ \times\widetilde{B}_z(v^\prime,w^\prime)\ ,
\end{split}
\end{equation}
with the kernels $K_{v,w}$ given in Appendix~\ref{app_kernel_Airr}. The boundaries $w=0$ and $w=W$ in the conformal space have to be chosen such that this volume includes the rim $\partial\Sigma$, i.e., it must map back to the 2D volume $\mathbb{R}^2\backslash \Sigma^-$. This result can then be transformed back into real space by inverting the conformal map.

Since the conformal map can be computationally expensive for numerical calculations, especially if one intends to sample different magnetic field configurations for the same device geometry, we point out that there is a convenient simplification that allows to remain in real space. For this purpose, note that we do not actually need the fully spatially resolved $\mathbf{A}_\text{irr}$, but only the integral over the bridge path, see Eq.~\eqref{eq:irr_gauge_RD} and subsequent discussion. Let us now suppose that the bridge path is a straight line in conformal space, perpendicular to the superconducting surfaces (i.e., $w$ goes from 0 to $W$, but $v$ stays fixed). Then, the path follows a longitudinal curve in real space $\mathcal{L}_v$ for fixed $v$ (see inset in Fig.~\ref{fig4_conformal}), always guaranteed to be perpendicular to $\partial\Sigma$. We can then compute the flux that is threaded through slices separated by nearby curves $\mathcal{L}_v$ and $\mathcal{L}_{v+\Delta v}$. In the continuum limit $\Delta v\rightarrow 0$, the slice has the infinitesimal area $d\mathcal{A}_v$. The phase drop for a junction $\nu$ at position $v$ can then be written as
\begin{equation}\label{eq_phi_conformal}
    \phi_\nu(v)=\frac{2\pi}{\Phi_{0}}\int^{V}_{0}\frac{dv^{\prime}}{V}\int^{v}_{v^{\prime}}dv^{\prime\prime}\iint_{d\mathcal{A}_{v^{\prime\prime}}}d^2r\widetilde{B}_{z}\ .
\end{equation}
This equation is a significant generalisation to a closed expression for $\phi_\nu$ found in Ref.~\cite{riwar_2022}. We further note that the above conformal approach can also be generalised to 3D (where the conformal space changes from cylinder to torus). A 3D generalisation can even be relevant for thin films, in case the applied magnetic field has a strong in-plane component (contrary to the case illustrated in Fig.~\ref{fig3_surface_charges}), where surface charges accumulate on the top and bottom areas. We reserve such considerations for future work, see also the Discussion section below.

Equation~\eqref{eq_phi_conformal} is particularly suitable for concrete numerical evaluations, where one can discretise the real and conformal spaces, and precompute `masks' that select out conformal slices as shown in the inset of Fig.~\ref{fig4_conformal}. These slices stay the same for a given geometry, and thus can be reused upon changing the input magnetic field distribution, rendering the implementation of Eq.~\eqref{eq_phi_conformal} efficient.
Note that for the device geometry with rectangular capacitor wings shown in Fig.~\ref{fig4_conformal}, if the junction is placed inside the region C (between the two superconducting wings), a straight line in conformal space is also a straight line in real space. Consequently, Eq.~\eqref{eq_phi_conformal} can be used for straight junction bridges as in Fig.~\ref{fig2_btilde}(a).

In fact, for rectangular capacitor wings the total conformal map can be provided by piecewise maps for the individual regions C (central), L (left), R (right), and F (far side). Within C, the conformal map is just unity, that is $\mathbf{r}=\mathbf{u}$. For the left and right side (L and R), we use the standard Schwarz-Christoffel mapping with the four corner points $v_{1,2,3,4}$ acting as the vertices of the corresponding polygon. For region L, the function describing this mapping is given as
\begin{equation}\label{eq_conformal_map}
    f(z)=2\frac{F\left(\arcsin\left[\frac{\sqrt{v_{2}-v_{4}}}{\sqrt{v_{1}-v_{4}}}\frac{\sqrt{z-v_{1}}}{\sqrt{z-v_{2}}}\right]\left|\frac{v_{2}-v_{3}}{v_{1}-v_{3}}\frac{v_{1}-v_{4}}{v_{2}-v_{4}}\right)\right.}{\sqrt{v_{1}-v_{3}}\sqrt{z-v_{4}}}\ ,
\end{equation}
where in the above formula $v_{1,2,3,4}$ are complex numbers with the imaginary (real) parts corresponding to the $-x$ ($y$) positions of the respective vertices, and $z=-ix+y$. The right region R follows from the same mapping, but mirrored to the other side of the device. The $x$-coordinate of the far side F is mapped onto itself up to a shift, whereas the $y$-component is stretched such that horizontal lines in the conformal space (fixed $w$) are continuous. This stretching factor is extracted from Eq.~\eqref{eq_conformal_map}. As a consequence, the two separate, semi-infinite far field regions in real space merge into a single finite region in conformal space. This conformal compactification is similar in spirit to other widely known mappings from (semi) infinite to finite coordinate systems, such as, e.g., stereographic projections.

The above solutions fulfill by construction $\int_0^v A_w =0$, and thus the requirement that the sum of surface charges vanishes (iii). Note, however, that Eq.~\eqref{eq_Airr_conformal} is only consistent and well-defined if $\iint d^2r \widetilde{B}_z =0$ (when mapping back from conformal to real space). If $\widetilde{B}_z$ did not integrate to zero, it would not be possible to find a single-valued solution for $\mathcal{A}_\text{irr}$ on the periodic conformal strip (cylinder) in Fig.~\ref{fig4_conformal}. Due to the holographic property of Eq.~\eqref{eq_holographic}, we know that this condition is satisfied if, equally, $\iint d^2r B_z =0$, consistent with Maxwell's equations. It is of importance for practical purposes, and poses in particular a challenge for efficient numerical calculations. Namely, if we choose a finite `simulation box' in real space on which we perform calculations [relevant especially for implementations of Eq.~\eqref{eq_phi_conformal}], we have to choose the box big enough, such that the magnetic flux neglected outside the box is sufficiently small not to have a quantitative impact on the results. This is particularly challenging for the computation of far-field sources, like flux lines, as we discuss further below. In order to render $\mathbf{A}_\text{irr}$ well-defined, and obtain smooth numeric results, we resort to a further trick: we separately compute the magnetic flux `lost' outside the box, and subtract it from the chosen input field, such as to enforce Gauss' law for magnetism on the finite box. As long as the lost flux is sufficiently small, the resulting output can be trusted.

\section{Magnetic dipoles}\label{sec_magnetic_dipoles}

As detailed in the introduction, the best available candidate model to describe the observed universality of $1/f$ flux noise is that of surface spins. We therefore consider ensembles of magnetic dipole moments located close to the superconducting surfaces. In addition to surface or substrate spins, the magnetic dipole solution is also valid for the far field of current-carrying flux lines (for flux control), which we will discuss further below. 

As indicated above, for thin films the relevant magnetic field is the $z$-component at $z=0$. For a dipole moment $\mathbf{m}=(m_x,m_y,m_z)$ located at $\mathbf{r}_0$, we have the standard expression (in 3D space)
\begin{equation}\label{eq_B_3D}
\mathbf{B}^{(0)}\left(\mathbf{r}\right)=\frac{\mu_{0}}{4\pi}\left[3\frac{\delta\mathbf{r}\left(\delta\mathbf{r}\cdot\mathbf{m}\right)}{\left|\delta\mathbf{r}\right|^{5}}-\frac{\mathbf{m}}{\left|\delta\mathbf{r}\right|^{3}}\right]\ ,
\end{equation}
with $\delta\mathbf{r}=\mathbf{r}-\mathbf{r}_0$. When projected onto the $z$-component at $z=0$, this reduces to
\begin{equation}\label{eq_Bz_2D}
    B_{z}^{(0)}\left(\mathbf{r}\right)=-\frac{\mu_{0}}{4\pi}\frac{m_{z}}{\left|\mathbf{r}-\mathbf{r}_0\right|^{3}}\ ,
\end{equation}
where here, both $\mathbf{r}$ and $\mathbf{r}_0$ are again in the 2D plane. Note that the area integral of Eq.~\eqref{eq_Bz_2D} is ill-defined, if the area includes the origin of the dipole. This divergence can be cured by pivoting the dipole moment ever so slightly out of the $z=0$ plane, and then integrating first, before going to the $z\rightarrow 0$ limit. In all results that follow, this procedure is performed implicitly (whenever necessary).

Note that while Eqs.~\eqref{eq_B_3D} and~\eqref{eq_Bz_2D} are commonly known as the magnetic field due to a classical dipole moment, we here assume the same magnetic field to be emitted also for quantum spins via the replacement $\mathbf{m}\rightarrow -g \mu_B \mathbf{\sigma}/2$, with the usual Landé $g$-factor, Bohr magneton $\mu_B$, and the vector of Pauli matrices $\mathbf{\sigma}=(\sigma_x,\sigma_y,\sigma_z)$. We note that indeed, the above form can also be derived when starting from a relativistic description of massive electrons with spin, see the Breit equation~\cite{Breit_1929}.

We include ensembles of fluctuating magnetic dipole moments $\widehat{\textbf{m}}_j$ by treating the moment as an operator (hat notation) and by introducing the index $j$, enumerating the individual spins/moments located at $\mathbf{r}_{j}$. Note that throughout this work, we will stick to 2D structures (in the $x,y$-plane), which couple dominantly to the $z$ component, $\widehat{m}_{z,j}$. An outlook on in-plane dipole orientations is, as already indicated above, given in the Discussion section.

The phase drop operator at junction $\nu$ can be written as
\begin{equation}\label{eq_phi_nu_hat}
    \widehat{\phi}_\nu(t)=\sum_j \alpha_{\nu}(\mathbf{r}_j)\widehat{m}_{z,j}\ ,
\end{equation}
where the coefficients $\alpha_{\nu,j}$ contain all the relevant information about the device geometry. For instance, for spin 1/2, we relate the magnetic moment to the spin via $\widehat{m}_j=-g\mu_B \widehat{\sigma}_j/2$ ($\hbar=1$). Plugging the above into the PSD of Eq.~\eqref{eq:PSD}, and going to a continuum representation, we get
\begin{equation}\label{eq_S_nu_nup}
\begin{split}
    S_{\nu,\nu^{\prime}}\left(\omega\right)=\frac{\left(\sigma g\mu_{B}\right)^{2}}{4}\int d^{2}r\int d^{2}r^{\prime}\alpha_{\nu}\left(\mathbf{r}\right)\alpha_{\nu^{\prime}}\left(\mathbf{r}^{\prime}\right)\\ \times S_{\sigma}\left(\mathbf{r},\mathbf{r}^{\prime},\omega\right)\ ,
\end{split}
\end{equation}
with the continuum spin density $\sigma=\sum_{j}\delta\left(\mathbf{r}-\mathbf{r}_{j}\right)$ (number of spins per area) and the spin-spin correlator
\begin{equation}\label{eq_S_spin_spin}
    S_{\sigma}\left(\mathbf{r},\mathbf{r}^{\prime},\omega\right)=\int^{\infty}_{-\infty}\mathrm{d}t\frac{e^{i\omega t}}{2}\left\langle \left\{ \sigma_{z}\left(t,\mathbf{r}\right),\sigma_{z}\left(0,\mathbf{r}^{\prime}\right)\right\} \right\rangle \ .
\end{equation}
For the spin-spin correlator $S_{\sigma}\left(\mathbf{r},\mathbf{r}^{\prime},\omega\right)$ we will resort to common shapes known from the spin diffusion model and generalisations thereof, see Refs.~\cite{Faoro_2008,Anton_2013,Lanting_2014,LaForest_2015,Aquino_2022}.

\section{Limitations of reciprocity}\label{sec_limitations_reciprocity}

We can contrast the above framework to the principle of reciprocity invoked by Refs.~\cite{Bialczak_2007,Lanting_2014,LaForest_2015,Aquino_2022}. These references work with the following expression to describe the coupling between a dc SQUID (with two junctions $\nu=1,2$) and a magnetic moment,
\begin{equation}\label{eq_reciprocity}
    \phi_1-\phi_2=\frac{2\pi}{\Phi_0}\frac{\mathbf{B}_{I,\text{loop}}(\mathbf{r}_j)\cdot \mathbf{m}_j}{I_\text{loop}} \ ,
\end{equation}
where $\mathbf{B}_{I,\text{loop}}(\mathbf{r}_j)$ is the magnetic field induced by a supercurrent of magnitude $I_\text{loop}$ circulating along the SQUID loop, measured at the position $\mathbf{r}_j$ of a magnetic moment $\mathbf{m}_j$. To include ensembles, one simply sums over the index $j$, similar to Eq.~\eqref{eq_phi_nu_hat}. The division by $I_\text{loop}$ makes sure that Eq.~\eqref{eq_reciprocity} is independent of the loop current magnitude, and that the formula has the correct units.

\begin{figure*}
    \centering
    \includegraphics[width=0.8\linewidth]{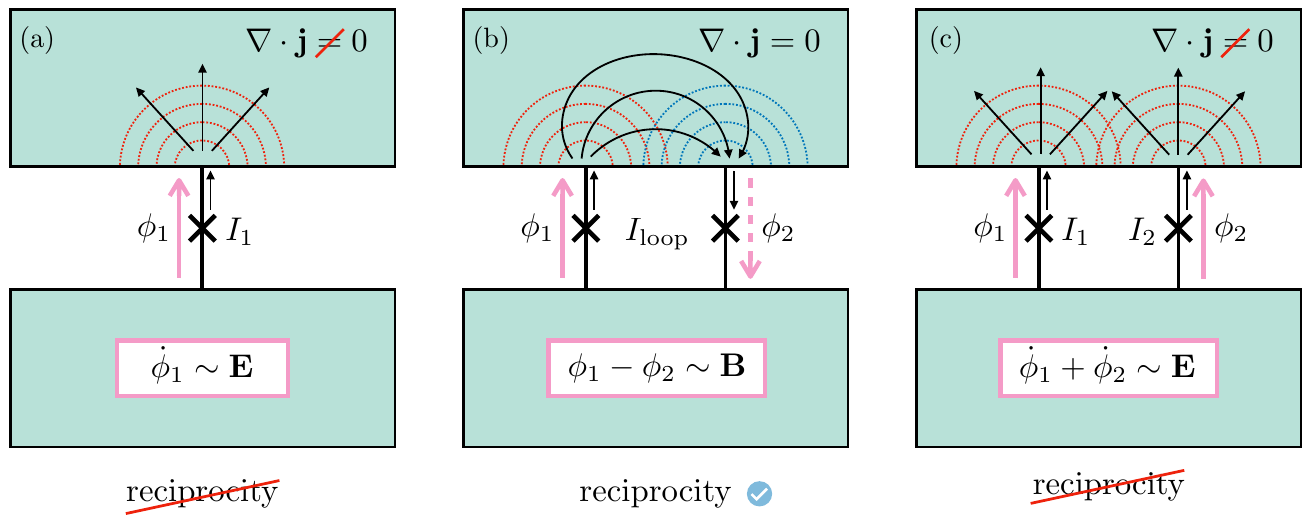}
    \caption{Central issue with the reciprocity principle to account for the Faraday effect. Reciprocity relies on computing the magnetic field induced by a supercurrent circulating within the device, which is then seen by a given magnetic dipole. For a single junction device (a) no such loop current exists. One can define a dipole current of magnitude $I_1$ going through junction 1, and compute its corresponding magnetic field, but this current is not source-free, making it impossible to generalise the reciprocity approach in this simple manner to capture the EMF mode. For a two-junction device we can define a loop current of magnitude $I_\text{loop}$ (b) and two dipole currents along individual junctions $I_1$ and $I_2$ (c). For the former (b) the standard reciprocity principle applies, allowing to compute $\phi_1-\phi_2$ by means of the magnetic field emitted by the loop current. For the latter (c) it does not for the same reason as in (a). }
    \label{fig5_reciprocity}
\end{figure*}

The above relation originates from general considerations, and indeed we have not found any instance where our formalism exhibits a contradiction with Eq.~\eqref{eq_reciprocity}. But we stress that the reciprocity relation only provides information on the flux mode $\phi_1-\phi_2$. It is therefore natural to wonder whether there exists a possible generalization or modification of the above formula for the EMF mode ($\phi_\nu$, such that $\dot{\phi}_\nu$ is proportional to the electromotive field across junction $\nu$, see also Fig.~\ref{fig1_main}). As we now outline, such a relationship is challenging for two reasons. 

First of all, both the left and right hand side of Eq.~\eqref{eq_reciprocity} are trivially gauge-invariant. Any effort to generalise Eq.~\eqref{eq_reciprocity} such that it captures the EMF mode have to grapple with gauge transformations, the central topic of this work. Let us cast this issue aside for a moment, and attempt a naive generalisation as follows. Let us assume that instead of a loop current we consider a `dipole' current leaving one of the transmon contacts (e.g., the lower wing) and enter the other (upper wing). Through a given junction $\nu$, we apply a current of fixed magnitude $I_\nu$, and postulate the formula
\begin{equation}\label{eq_reciprocity_nu}
    \phi_\nu = \frac{2\pi}{\Phi_0}\frac{\mathbf{B}_{I,\nu}(\mathbf{r}_j)\cdot \mathbf{m}_j}{I_\nu}\ . 
\end{equation}
The obvious problem with this construction is that -- contrary to the loop current -- the dipole current cannot possibly be source-free as charges accumulate in the capacitor wings, see Fig.~\ref{fig5_reciprocity}. Consequently, it is in general unclear, how exactly to define $\mathbf{B}_{I,\nu}$.

This would render a generalisation from Eq.~\eqref{eq_reciprocity} to Eq.\eqref{eq_reciprocity_nu} impossible in this simple manner. But there is one exception. Let us allow for our model to have capacitor wings of infinite size. In that case, a source-free dipole current can flow, since the charge flowing into (out of) the transmon wings is dissipated into an infinite volume. In what remains of the Results section, we consider two cases, one of nominally infinite capacitor wings and a second one where finite capacitances are required. And indeed, in the former case, we can explicitly show that the naive generalisation Eq.~\eqref{eq_reciprocity_nu}, surprisingly, holds. Below, we derive all results by means of our irrotational framework, not relying on this generalised notion of reciprocity. But we will provide the equivalent dipole current density distribution (whenever it exists) which determines $\mathbf{B}_{I,\nu}$.

Equation~\eqref{eq_reciprocity_nu} is first of all informative on a conceptual level. Our (irrotational) gauge-fixing procedure for $\phi_\nu$ allows, at least for the special case of infinitely large capacitor wings, to connect a nominally gauge-dependent quantity ($\phi_\nu$) to a physically transparent process in the form of the (gauge-invariant) magnetic field emitted due to a constant dipole current flowing between the transmon wings across junction $\nu$, $\mathbf{B}_{I,\nu}$. Secondly, we expect this shortcut to be of use for follow-up research. Namely, one could benefit from the increased simplicity of Eq.~\eqref{eq_reciprocity_nu} for more complicated device designs, such as junction bridges that are not straight lines, where the simplified Eq.~\eqref{eq_phi_conformal} is no longer valid.

We nonetheless reemphasise that the generalised reciprocal formula, Eq.~\eqref{eq_reciprocity_nu}, only holds for infinite capacitor wings. In the last part of the Results section, we present calculations where it does not hold. Overall, this means that reciprocity, at least as formulated above, is of limited use in order to properly account for the Faraday effect. We cannot exclude the existence of a more sophisticated generalisation capable of correctly including charge displacements relevant for finite capacitor wings -- a further possible subject of follow-up research. However, it may well be that such a generalisation effort undoes the conceptual advantage gained by the reciprocity principle in the first place. For now, we maintain that the irrotational framework presented here, notably Eqs.~\eqref{eq_Airr_conformal} as well as the special case of Eq.~\eqref{eq_phi_conformal}, represent a generally valid recipe to account for the coupling between magnetic fields and superconducting thin films, correctly including Faraday's law.

\section{Junction bridges versus capacitor wings}\label{sec_brigdes_vs_wings}

We now apply the above formalism for a transmon geometry with infinitely large capacitor wings, relevant for magnetic dipoles located close to the junction bridges. Furthermore, we neglect the Meissner effect, which turns out to be an accurate approximation (as demonstrated explicitly in Appendix~\ref{app_Meissner_irrelevant}). Under these simplifying assumptions, we arrive at an analytic, closed expression for $\phi_\nu$, and thus make precise quantitative statements about the impact of the Faraday effect on the qubit performance. In particular, we will discuss under what assumptions the upper limit for the quality factor [Eq.~\eqref{eq_Q_limit}] due to universal flux noise holds, and will identify a regime where this limit might even suffer a further reduction.

\begin{figure*}
    \centering
    \includegraphics[width=0.8\linewidth]{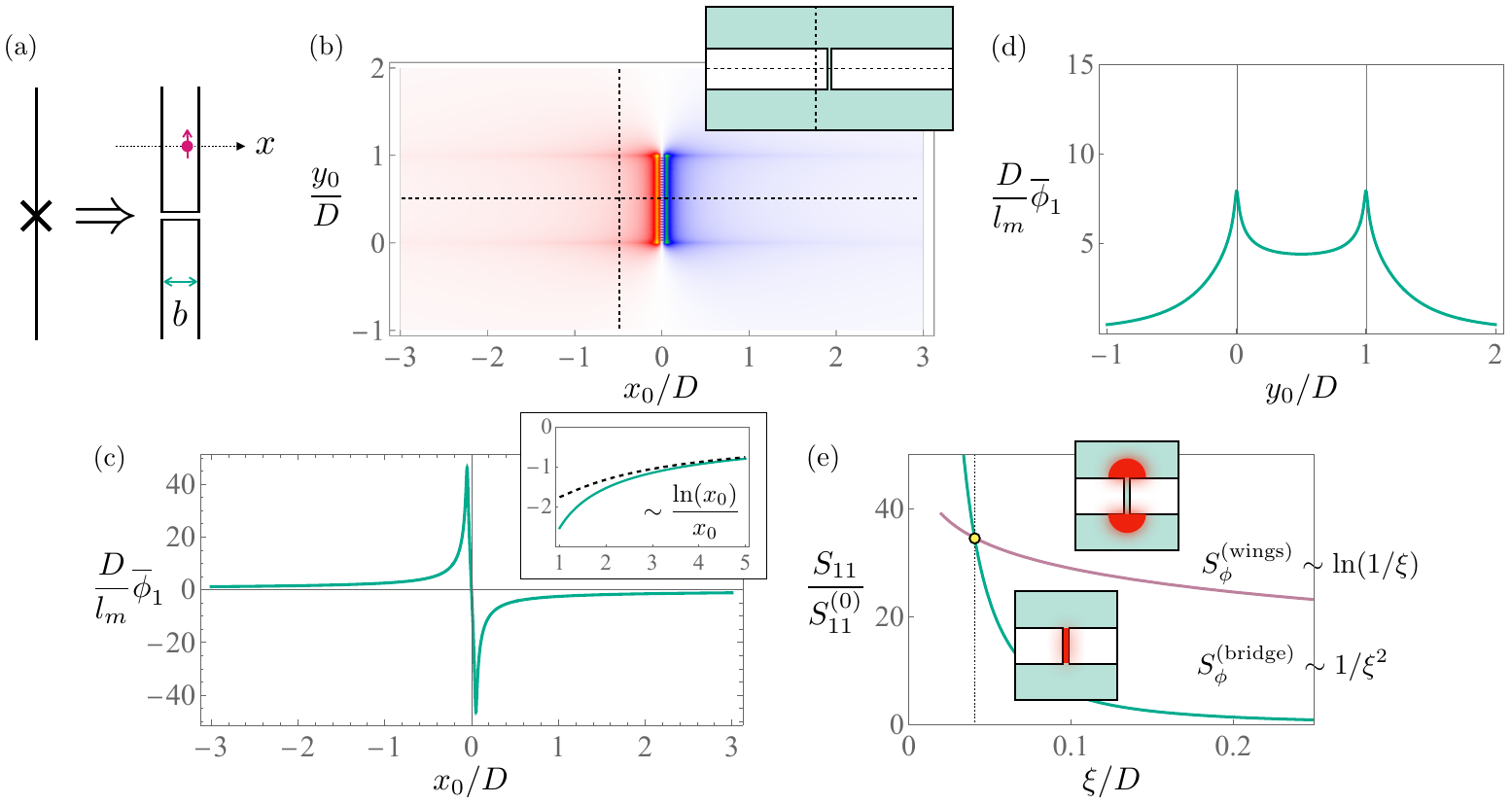}
    \caption{Interaction of a spin with a single-junction (fixed-frequency) transmon depending on the spin position. (a) In order to regularise the divergence of the coupling to the junction bridge, we introduce a finite bridge width $b$. (b) Density plot of $\overline{\phi}_1$ as a function of the spin position $x_0$ and $y_0$. Length units are taken with respect to the bridge length $D$. The inset shows the superconducting thin film structure (green area). (c) and (d) are cross section plots at the horizontal and vertical dashed lines indicated in (b). The phase drop $\overline{\phi}_1$ experiences logarithmic divergencies at the bridge edges, and a sign change when going from one edge to the other (c). The capacitor wing edges also provide a logarithmic divergence, but no sign change between the upper and lower wing (d). In (b-d), the remaining parameters are $b=0.1D$ and $\delta z=0.01 D$. (e) By introducing a finite spin correlation length $\xi$, we can show a crossover from a bridge-dominated to a wing-dominated qubit relaxation rate at $b\approx \xi$. In (e) $b=0.05D$. }
    \label{fig6_bridge_vs_wing}
\end{figure*}

For concreteness we fix the coordinate system as follows, see also Fig.~\ref{fig6_bridge_vs_wing}(b). The edge of the upper (lower) capacitor wing is located at $y=D$ ($y=0$). The bridge containing the Josephson junction shall be at $x=0$. Since we assume capacitor wings of infinite size, we only have the central region C between the two wings to consider, such that the conformal mapping step illustrated in Fig.~\ref{fig4_conformal} is not necessary. 

As we neglect the Meissner effect, we use as the input simply the bare magnetic field of the dipole, i.e., we set $B_z=B_z^{(0)}$ and insert for $B_z^{(0)}$ the dipole field given in Eq.~\eqref{eq_Bz_2D}. Note that zero Meissner screening does not mean that screening supercurrents play no role in the considered problem. To the contrary, screening currents emerge due to the London equation (here formulated in 2D)
\begin{equation}
    \dot{\mathbf{J}}=\frac{2}{\mu_0\Lambda}\mathbf{E}\ ,
\end{equation}
which is yet again nonzero due to Faraday's law. Moreover, while $\mathbf{J}\sim 1/\Lambda$ (going to $0$ for $\Lambda\rightarrow\infty$) the screening current contribution in $\widetilde{B}_z$, Eq.~\eqref{eq_btilde}, enters with a prefactor $\Lambda$, cancelling the $1/\Lambda$ suppression of $\mathbf{J}$. As a consequence, the transversal electric field (due to screening currents) is actually of the same order of magnitude as the longitudinal field (which connects to surface charges due to Thomas-Fermi screening).

Consequently, we still need to compute the 2D vector potential $\mathbf{A}$ in the London gauge for the interior of the superconducting strips, which is here possible analytically. We can express $\mathbf{A}$ in terms of the stream function $\psi$, $\mathbf{A}=(-\partial_y\psi,\partial_x\psi)$, which automatically satisfies $\nabla \cdot\mathbf{A}=0$ by construction. This yields the Poisson equation for the stream function
\begin{equation}
    \nabla^2\psi(\mathbf{r})=B_z(\mathbf{r})\ .
\end{equation}
In order for $\mathbf{A}$ to be source-free also at the boundaries of the capacitor wings (at $y=0,D$), we need to impose in addition $\psi=0$ at those boundaries. For the upper wing ($y\geq D$), this boundary condition is included in the solution of the 2D Poisson equation via
\begin{equation}
    \psi\left(\mathbf{r}\right)=\int_{-\infty}^{\infty}dx^{\prime}\int_{D}^{\infty}dy^{\prime}S(x-x^\prime,y,y^\prime)B_{z}\left(x^{\prime},y^{\prime}\right)\ 
\end{equation}
with the kernel
\begin{equation}
    S(x,y,y^\prime)=\frac{1}{2\pi}\ln\left(\frac{\sqrt{x^{2}+\left(y-y^{\prime}\right)^{2}}}{\sqrt{x^{2}+\left(y+y^{\prime}-2D\right)^{2}}}\right)\ .
\end{equation}
The ultimate quantity of interest is the vector potential parallel to the edge, $A_{x}\left(x,0\right)=\left.-\partial_{y}\psi\right|_{y=0}$, which gives,
\begin{equation}\label{eq_Ax_edge_London}
    A_x\left(x,0\right)=\frac{1}{\pi}\int dx^{\prime}\int_{0}^{\infty}dy^{\prime}\frac{y^{\prime}B_{z}\left(x^{\prime},y^{\prime}\right)}{\left(x-x^{\prime}\right)^{2}+\left(y^{\prime}\right)^{2}}\ .
\end{equation}
This is the boundary term that we need in order to fix the parallel (transversal) component of the irrotational gauge for the space outside the superconductor, in between the two capacitor plates, as per above condition (ii). The resulting (gauge invariant) screening current at the edge $x=0$ is thus given as
\begin{equation}\label{eq_jx_edge}
    J_x(x,0)=-\frac{2}{\mu_0\Lambda}A_x(x,0)\ ,
\end{equation}
with $A_x$ given by Eq.~\eqref{eq_Ax_edge_London}. This term is now introduced into the construction of $\widetilde{B}_z$ as given in Eq.~\eqref{eq_btilde}. The lower capacitor with edge at $y=0$ provides an additional contribution of the same type in a mirrored fashion, where one has to account for an overall sign change, since the normal vector $\mathbf{n}$ to the upper (lower) edge points downwards (upwards).

The resulting $\widetilde{B}_z$ field is now inserted into Eq.~\eqref{eq_phi_conformal}, with the simplification that since we consider only the central area C between the wings, we can replace conformal coordinates $(v,w)$ with regular position coordinates $(x,y)$. 
We thus find that for a dipole at $(x_0,y_0)$, Eq.~\eqref{eq_Bz_2D}, the phase drop across a junction $\nu$ at position $x=x_\nu$ (going in a straight line from $y=0$ to $y=D$)
yields 
\begin{equation}\label{eq_phi_2D_spin}
\begin{split}
    \phi_\nu(x_\nu)=\frac{l_m}{2}\sum_{y\in\text{edges} }h^{\prime}\left( \delta x_{\nu},y\right)\left[\frac{\delta x_\nu y}{\delta x_{\nu}^2+\delta z^2}\right.\\ \left.\phantom{\frac{y}{x_{\nu}}}+\frac{1}{\pi}h\left(\delta x_{\nu},y\right)\right]\ ,
\end{split}
\end{equation}
where $\delta x_\nu=x_\nu-x_0$ is the distance between bridge and dipole.
The notation $y\in \text{edges}$ means that we sum $y$ over the two values $y=y_0$ and $y=D-y_0$ (distance of the dipole from the two capacitor wing edges). The auxiliary function is
\begin{equation}
    h(x,y)=\ln\left(\frac{\sqrt{x^2+y^2+\delta z^2}+x}{\sqrt{x^2+y^2+\delta z^2}-x}\right)\ ,
\end{equation}
and $h^\prime=\partial_x h$. We kept a finite distance $\delta z$ of the dipole from the plane $z=0$ to smooth out the function. We furthermore introduced the magnetic length scale associated with the dipole moment $l_m=\mu_0 m_z/(2\Phi_0)$.

Note that $\phi_\nu$ exhibits a $1/r$ divergence close to the bridge for $\delta z\rightarrow 0$, which can be regularised by including a finite bridge width $b$, Fig.~\ref{fig6_bridge_vs_wing}(a), as already foreshadowed in the discussion following Eq.~\eqref{eq:irr_gauge_RD}. This is conveniently accomplished by averaging over bridge positions,
\begin{equation}\label{eq_phi_nu_regularisation_b}
    \overline{\phi}_\nu=\frac{1}{b}\int_{x_\nu-\frac{b}{2}}^{x_\nu+\frac{b}{2}}dx \phi_\nu(x)\ .
\end{equation}
The heat map in Fig.~\ref{fig6_bridge_vs_wing}(b) shows $\overline{\phi}_1$ for a single junction positioned at $x=0$ as a function of the dipole position, indicating both the magnitude and sign (red versus blue) of the coupling. Figures~\ref{fig6_bridge_vs_wing}(c) and~(d) show cross sections cutting through either the bridge or the capacitor wings, respectively.

Close to the bridge, one obtains logarithmic divergencies,
\begin{equation}\label{eq_phi1_bridge}
    \frac{\overline{\phi}_1}{l_m}\approx \frac{2}{b}\ln\left(\left\vert\frac{b/2-x_0}{-b/2-x_0}\right\vert\right)\ .
\end{equation}
A logarithmic divergence is also present at the capacitor wing edges, e.g., for the lower wing $\overline{\phi}_1 \sim \ln(1/y_0)/x_0$. However, since $b$ is the smallest length scale in the problem, the bridge divergence is nominally much stronger than the capacitor contribution, see also Figs.~\ref{fig6_bridge_vs_wing}(b-d).

But note that depending on what dipole distribution we assume, we have to integrate over entire parts of the device in order to compute the resulting relaxation rate, Eq.~\eqref{eq_Gamma_1}. That is, we do not have to consider local coupling strength, but the one integrated over extended areas. Moreover, literature has proposed various spin models~\cite{Faoro_2008,Anton_2013,Lanting_2014,LaForest_2015,Aquino_2022}, such as the spin diffusion model~\cite{Faoro_2008}, which exhibit a finite correlation length $\xi$, which provides a further interesting complication as we show now. 


For a single junction, the relaxation rate of Eq.~\eqref{eq_Gamma_1} is directly proportional to the correlation function $S_{11}$, Eq.~\eqref{eq_S_spin_spin}, when inserting above result for $\phi_1$. We assume, in accordance with~\cite{Faoro_2008,LaForest_2015}, that the dominant contribution comes from spins located on the superconducting surfaces. As a consequence, we have to integrate over the bridge ($x$ from $-b/2$ to $+b/2$ and $y$ from $0$ to $D$) and the capacitor wing ($y\geq D$ and $y\leq 0$) areas. We compute bridge and wing contributions to the relaxation rate separately. As we will explain below, this separation is important for the appropriate estimation of the quality factor limit postulated in Eq.~\eqref{eq_Q_limit}.

As for the spin-spin correlation function, Eq.~\eqref{eq_S_spin_spin}, there is always a sufficiently coarse-grained level (considering length scales larger than $\xi$), where correlations are local, $S_\sigma(\mathbf{r},\mathbf{r}^\prime)\approx s_\sigma \delta(\mathbf{r}-\mathbf{r}^\prime)$, where $s_\sigma$ has units of area times time. For the bridge, we allow $\xi$ to be comparable to $b$ (while $D$ always larger than $\xi$) and analyse the transition from $b\gg \xi$ to $b\ll\xi$. The $\delta$-correlation hypothesis therefore has to be relaxed for the $x$ dimension, which we do as follows. For the standard spin diffusion model~\cite{Faoro_2008}, the correlator is of the form $1/(1+k^2\xi^2)$ in Fourier space. Transforming back to real space (for the remaining $x$-direction), we arrive at
\begin{equation}
    \int_0^D dy \int_0^D dy^\prime S_\sigma(\mathbf{r}-\mathbf{r}^\prime)\approx  \frac{Ds_\sigma}{2\xi}e^{-\frac{\vert x-x^\prime \vert}{\xi}}\ .
\end{equation}
Using the simplified expression of Eq.~\eqref{eq_phi1_bridge} (valid close to the bridge), we get
\begin{equation}
\begin{split}
    S_{11}^\text{bridge}\approx S^{\left(0\right)}_{11}\frac{2D}{\xi b^{2}}\int_{-\frac{b}{2}}^{\frac{b}{2}}dx\int_{-\frac{b}{2}}^{\frac{b}{2}}dx^\prime \ln\left(\left\vert\frac{\frac{b}{2}-x}{-\frac{b}{2}-x}\right\vert\right)\\
    \times \ln\left(\left\vert\frac{\frac{b}{2}-x^\prime}{-\frac{b}{2}-x^\prime}\right\vert\right)e^{-\frac{\vert x-x^\prime \vert}{\xi}}\ ,
\end{split}
\end{equation}
where $S_{11}^{(0)}=\sigma^2 l_m^2s_{\sigma}$ captures the spin specific prefactors. Note that for spins, the magnetic length $l_m$ requires inserting $m_z\rightarrow g\mu_B/2$ for the magnetic moment. The term following $S_{11}^{(0)}$ is dimensionless and captures the information of the device geometry. For $\xi \ll b$, we again go to the local limit, arriving at
\begin{equation}\label{eq_S_11_bridge_local}
    \frac{S_{11}^\text{bridge}}{S_{11}^{(0)}}\approx \frac{4\pi^2D}{3b}\approx 13.15\frac{D}{b}\ .
\end{equation}
However, due to the combination of logarithmic divergences and the sign change across $x$, see Fig.~\ref{fig6_bridge_vs_wing}(c), this result is highly fragile. For $\xi\gtrsim b$, we enter a regime of a suppressed relaxation rate,
\begin{equation}\label{eq_S_11_bridge_xi}
    \frac{S_{11}^\text{bridge}}{S_{11}^{(0)}}\approx \frac{2}{9}\left(15-\pi^2\right)\frac{Db}{\xi^2}\approx 1.14\frac{Db}{\xi^2}\ .
\end{equation}
The correlator $S_{11}$ thus suffers a combined orders of magnitude reduction due to the numerical prefactor (stemming from the fragile logarithmic divergences) and a quadratic reduction $\sim b^2/\xi^2$ due to the finite coherence length (averaging over positive and negative coupling areas).

Let us compare this with the capacitor wing contribution. Since we here consider length scales larger than $b$, we here do not average over the bridge width, thus using $\phi_1$ instead of $\overline{\phi}_1$. Since we integrate over the entire wing plate, we use the long range approximation, as shown in the inset of Fig.~\ref{fig6_bridge_vs_wing}(c). For example, for spins positioned on the lower wing ($y_0<0$) we get 
\begin{equation}
    \frac{\phi_1}{l_m}\approx \frac{1}{\pi}\frac{\ln\left(\frac{\sqrt{x_0^2+y_0^2}+x_0}{\sqrt{x_0^2+y_0^2}-x_0}\right)}{\sqrt{x_0^2+y_0^2}}\ .
\end{equation}
We insert this into $S_{11}$, Eq.~\eqref{eq_S_nu_nup}, using the coarse-grained approximation for the spin-spin correlator, $S_\sigma(\mathbf{r},\mathbf{r}^\prime)\approx s_\sigma\delta(\mathbf{r}-\mathbf{r}^\prime)$ with a caveat: the above formula yields a $1/r$-divergence in the integrand of $S_{11}$ at the point where the junction bridge attaches to the wing, requiring a regularization. We again use the information that at the bridge, the coupling changes sign rapidly. Hence, for $\xi>b$, the cutoff is $\xi$ itself. We get (summing over both wings)
\begin{equation}\label{eq_S_11_wing}
    \frac{S_{11}^\text{wing}}{S_{11}^{(0)}}\approx 2\pi \ln\left(\frac{\sqrt{\mathcal{A}_\text{wing}}}{\xi}\right)\ ,
\end{equation}
where $\mathcal{A}_\text{wing}$ is the capacitor wing area. Notice how the wing contribution, Eq.~\eqref{eq_S_11_wing}, is significantly less (logarithmically) sensitive to device geometry and correlation length $\xi$ as compared to the bridge contribution in either regime, Eqs.~\eqref{eq_S_11_bridge_local} and~\eqref{eq_S_11_bridge_xi}. In particular, when we compare the two contributions as a function of $\xi$, Fig.~\ref{fig6_bridge_vs_wing}(e), we see that the wing contribution becomes dominant as soon as $\xi\gtrsim b$.

\begin{figure}
    \centering
    \includegraphics[width=\linewidth]{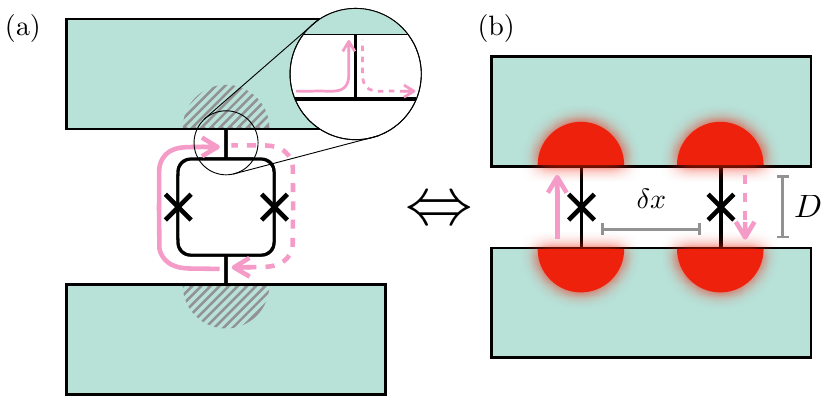}
    \caption{Difference between standard modern dc SQUID device geometries (a) and a hypothetical design (b) which could reveal a possibly dominant capacitor wing contribution. The reason why the capacitor wings do not enter the flux mode in design (a) is that the loop supercurrent flows mainly within the bridge filaments, whereas the supercurrent contribution into (or out of) the capacitor wing cancels (inset). In (b) such a cancellation is impossible.  }
    \label{fig7_wing_test}
\end{figure}

This information is important for the correct estimation of the quality factor limit, Eq.~\eqref{eq_Q_limit}. To this end, we combine two basic observations. First, note that the flux noise is measured by means of the dephasing rate, Eq.~\eqref{eq_Gamma_varphi}, which couples to the flux mode $\phi_1-\phi_2$. Second, usual dc SQUID devices have a loop formed from bridge parts only, see Fig.~\ref{fig7_wing_test}(a). For such a geometry, the flux mode can be represented (via reciprocity) with a supercurrent flowing along the loop, avoiding the capacitor wing portion of the device altogether. In contrast, for the EMF mode, $\phi_1+\phi_2$ (controlling the relaxation rate), the wing contribution cannot be avoided, because the corresponding equivalent supercurrent (dipole current) must flow into the capacitor wing, see also Fig.~\ref{fig5_reciprocity}. Hence, if $\xi \gtrsim b$ were satisfied, inferring the relaxation rate from flux noise amplitude measurements of the dephasing rate could lead to a severe underestimation of the former, perhaps even by an order of magnitude. Consequently, the fundamental limit of  $Q\sim 10^8\sim 10^{10}$ might suffer a further reduction because of this effect. Since the most recent devices are only beginning to approach the here obtained 
quality factor limit, we expect that a potential dominance of the capacitor wing contribution can as of right now only be tested indirectly. A possibility of such a test could be a dephasing measurement for an alternative dc SQUID design where the junction bridges do not converge before they join the capacitor wings, but connect separately, see Fig.~\ref{fig7_wing_test}(b). If the wing contributions are indeed dominant, the dephasing rate should worsen significantly.

Note that the above procedure strives to take results from actually measured dephasing rates, and use them to postulate a fundamental limit for the quality factor (relaxation rate). There remain a number of open challenges for more precise statements. The first is the frequency dependence which has so far been sidelined in the discussion. The spin diffusion model~\cite{Faoro_2008} exhibits a dynamic spin correlation length scaling as $\xi\sim 1/\sqrt{w}$. Given that the relevant frequency windows for the dephasing rate ($\text{mHz}\sim\text{Hz},\text{kHz}$) and the relaxation rate ($\text{GHz}$) differ by 6 orders of magnitude (or more), the relevant value of $\xi$ might vastly differ for the two rate types. More general classes yield more general power laws, $\xi\sim \omega^{-1/z}$ with a potentially reduced frequency sensitivity (e.g., $z=4$ for model B)~\cite{Hohenberg_review_1977}. We furthermore point to the possibility of a very weak logarithmic dependence of the correlation length on frequency due to thermally activated spin droplets~\cite{Fisher_1988}. Overall, given the reported stability of the $1/f$ noise law over wide frequency ranges (sub $\text{Hz}$ to low $\text{GHz}$)~\cite{Quintana_2017}, we expect a weak frequency-dependence of $\xi$ to be plausible, which would corroborate the validity of the above estimate.

A second challenge, suitable for follow-up research, concerns further details of the device geometry. For a theoretical modelling of the flux mode, Ref.~\cite{LaForest_2015} included also spins with orientations within the ($x,y$)-plane. While such spins couple much weaker to thin films (as already explained), they do not suffer from a reduction due to finite $\xi$, since the coupling strength has no sign change. The challenge to include in-plane magnetic dipoles for the EMF-mode is that they violate the condition illustrated in Fig.~\ref{fig3_surface_charges}. Since the magnetic field is no longer dominantly orthogonal to the plane, Thomas-Fermi screening charges now occur also at the top and bottom surfaces of the thin films, not only on the 1D rim. This requires a true 3D treatment of the Maxwell problem in spite of working with thin films.

Finally, throughout this work we assume the junction bridges to be straight lines (relevant for single junction transmons). In order to make an even better quantitative connection between dephasing and relaxation rates, we should include the bridge corners that form the typical dc SQUID loops, see Fig.~\ref{fig7_wing_test}(a). We note that this detail is most conveniently added by using the reciprocity relationship, either Eq.~\eqref{eq_reciprocity} for the flux mode, or Eq.~\eqref{eq_reciprocity_nu} for the EMF mode (valid here due to the infinite size capacitor wings).

In this context, we conclude this section by providing the explicit supercurrent distributions that produce the equivalent induced magnetic field for the reciprocity relation, Eq.~\eqref{eq_reciprocity_nu}. For the current distribution, we use
\begin{equation}
    \mathbf{j}(\mathbf{r})=I\delta\left(z\right)\left\{ \begin{array}{cc}
\left(\begin{array}{c}
0\\
1
\end{array}\right)\delta\left(x\right) & 0<y<D\\
\frac{1}{\pi}\left(\begin{array}{c}
\frac{x}{x^{2}+\left[y-D\right]^{2}}\\
\frac{y-D}{x^{2}+\left[y-D\right]^{2}}
\end{array}\right) & y>D\\
-\frac{1}{\pi}\left(\begin{array}{c}
\frac{x}{x^{2}+y^{2}}\\
\frac{y}{x^{2}+y^{2}}
\end{array}\right) & y<0
\end{array}\right.\ .
\end{equation}
As long as the wings are open ended, this distribution satisfies $\nabla\cdot\mathbf{j}=0$. Inserting it into Ampere's law, Eq.~\eqref{eq_ampere}, and inserting the resulting magnetic field into Eq.~\eqref{eq_reciprocity_nu}, we arrive at the result given in Eq.~\eqref{eq_phi_2D_spin}, establishing the equivalence between the irrotational and reciprocal approaches. The generalisation to a finite bridge width $b$ follows also here very simply by a convolution of the same type as in Eq.~\eqref{eq_phi_nu_regularisation_b}.

\begin{figure}
    \centering
    \includegraphics[width=\linewidth]{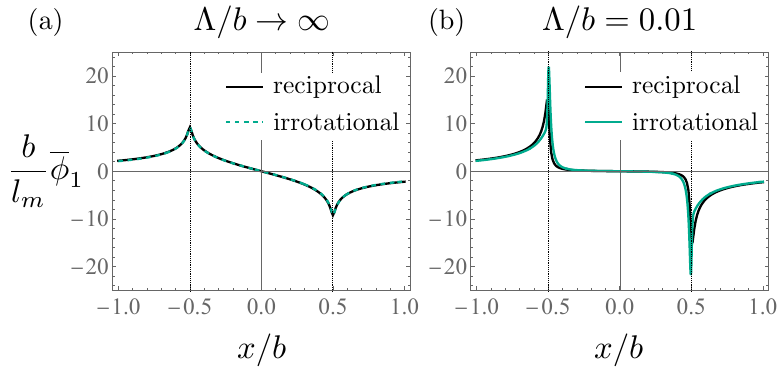}
    \caption{Comparison between the reciprocal and irrotational approach to compute the phase drop $\overline{\phi}_1$. We here assume an infinitely long bridge ($D\rightarrow \infty$), neglecting the capacitor wing contribution. Without the Meissner effect ($\Lambda\rightarrow \infty$) the agreement between both approaches is exact (a). In the presence of a finite Meissner effect the agreement is approximate (b), because the reciprocal approach neglects a backaction effect between spin and supercurrent, see also the main text.}
    \label{fig8_reciprocal_test}
\end{figure}

While we here neglected the Meissner effect, we note that the equivalence can also be demonstrated in the presence of it. We do so numerically, for simplicity close to the bridge, i.e., we neglect the capacitor wing contributions (setting $D\rightarrow\infty$ and considering $y$ in the proximity of $D/2$). In that case, one can efficiently resort to a quasi 1D representation (retaining the finite bridge width $b$). For the irrotational approach, we use the Meissner screening formula, Eq.~\eqref{eq_Meissner_2D}, with the resulting screened magnetic field entering Eq.~\eqref{eq_phi_conformal}. This can be compared to the result from the reciprocal formula, Eq.~\eqref{eq_reciprocity_nu}, when modifying the current distribution $\mathbf{j}$ to account for the Meissner effect, which can be done when including an inductive current-current coupling. The results for $\overline{\phi}_1$ are shown in Fig.~\ref{fig8_reciprocal_test} either with ($\Lambda$ finite) or without ($\Lambda\rightarrow \infty$) the Meissner effect. Note that in the presence of the Meissner effect, the reciprocal and irrotational results agree only approximately, Fig.~\ref{fig8_reciprocal_test}(b). This is because in the reciprocal approach, we only took into account the Meissner coupling of the supercurrent with itself, but not the backaction between supercurrent and dipole. This backaction effect has already been pointed out in Ref.~\cite{LaForest_2015} for the reciprocal calculation of the flux mode in superconducting wires. The authors of Ref.~\cite{LaForest_2015} point out that the backaction effect can be included by means of the FastHenry software package. We here note that the irrotational approach, in addition to being able to calculate the EMF mode, contains a slight conceptual advantage also for the flux mode. Namely, by starting right away with the full screened magnetic field, Eq.~\eqref{eq_Meissner_2D}, this backaction effect is automatically included without the need for further refinement.

\section{Safety distance for flux lines}\label{sec_safety_distance}

\begin{figure*}
    \centering
    \includegraphics[width=0.7\linewidth]{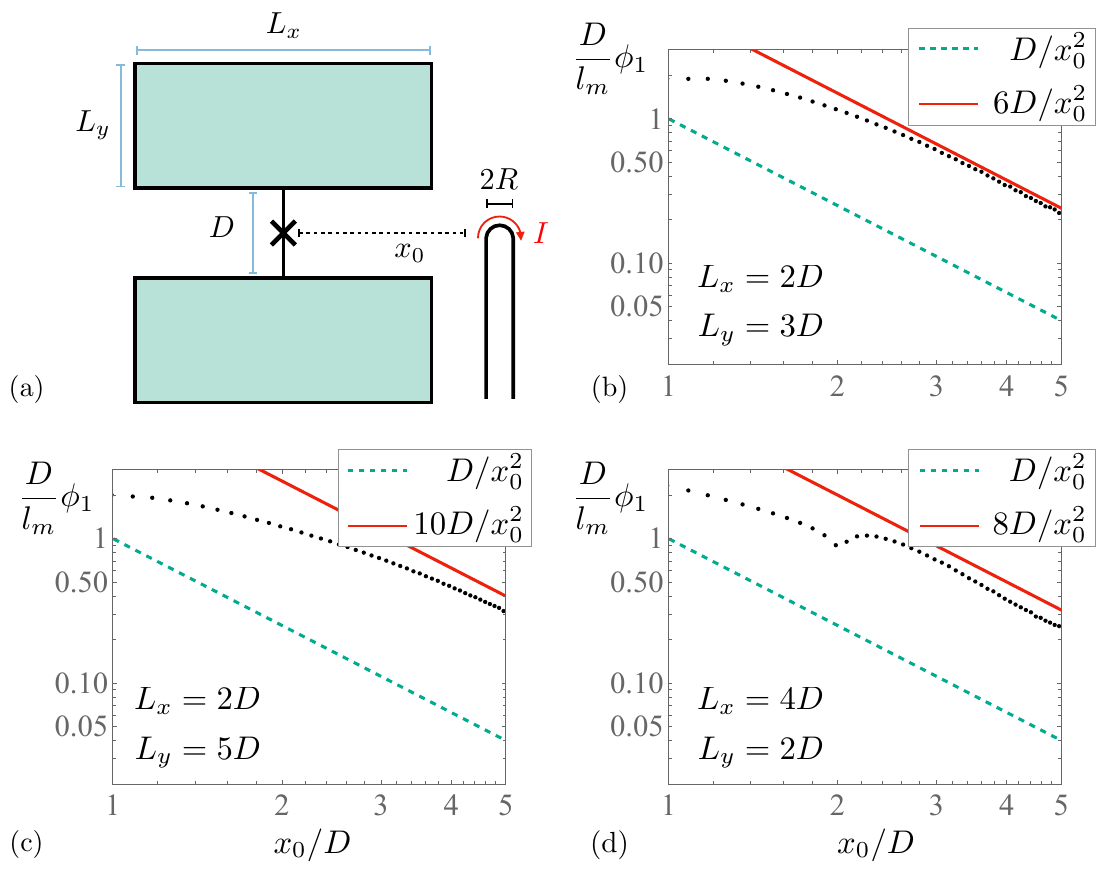}
    \caption{The coupling of a flux line to a single-junction transmon, placed at a distance $x_0$ from the junction. (a) The flux line consists of a collinear wire with a loop termination of radius $R$ at the tip. Upon sending current $I$ through the wire, it will create a magnetic dipole field with magnetic moment $m_z=\pi R^2 I$. (b-d) The phase drop across the junction $\phi_1$ as a function of the distance $x_0$, for different capacitor wing dimensions $L_x,L_y$. Without the wings, the EMF field should scale as $\phi_1/l_m\approx D/x_0^2$ (turquoise dashed lines). The numerical computation (black dotted) reveals that while the asymptotic behaviour indeed approaches $\sim 1/x_0^2$, there is a device geometry-dependent boosting factor roughly given as $L_xL_y/D^2$ (red). The latter is due to the holographic property of the relevant field $\widetilde{B}_z$ which contains the information of the flux piercing the capacitor wing interior, see also Fig.~\ref{fig2_btilde}. }
    \label{fig9_flux_lines}
\end{figure*}

We have so far focussed on surface spins, which naturally couple very locally to the circuit. We now want to discuss far field sources, where the magnetic field varies only weakly near the superconducting device. This is the case, for instance, for current-carrying flux lines, required for dc SQUID flux-control, see Fig.~\ref{fig9_flux_lines}(a). While such lines obviously couple to the flux mode of SQUIDs, we here focus on the EMF mode, which can, in the same spirit as the rest of this work, couple also to single junction devices.

Usually, far-field contributions can be analysed much more efficiently than near-field sources such as surface spins, because one can, in a first approach, assume a nearly constant magnetic field $B_z\approx \text{const.}$ near the device, not having to deal with the divergences we encountered above. However, note that this is only true for the flux mode. For the EMF mode, a constant magnetic field approximation cannot be successful, specifically because of the screening surface charges. This fact is best understood when invoking again the conformal map picture, see Fig.~\ref{fig4_conformal}. A constant field approximation would automatically provide a divergent contribution in the far region F, resulting in an ill-defined description of surface charges screening the electromotive field. In reality, the surface charges on the edge facing F are well-defined but nonetheless nonzero, and contribute noticeably to $\phi_1$. We therefore have to keep the capacitor wing size finite, rendering the previously explored generalisation of the reciprocity relation to the EMF mode, Eq.~\eqref{eq_reciprocity_nu}, invalid.

We therefore have to carefully apply the irrotational formalism keeping the full information of the magnetic field everywhere in free space. We do so by implementing Eq.~\eqref{eq_phi_conformal} numerically using discrete conformal slices, like the ones shown in the inset of Fig.~\ref{fig4_conformal}. As for the magnetic field source, we exploit the fact that the tips of the collinear flux lines shown in Fig.~\ref{fig9_flux_lines}(a) can be effectively modelled as simple current carrying rings of a given radius $R$ (since in the far-field limit, the current feed to and from the tip cancels). For distances larger than $R$, this field source shrinks again to an effectively 0-dimensional magnetic dipole, Eq.~\eqref{eq_Bz_2D}, with magnetic moment $m_z=\pi I R^2$. We can therefore continue working with Eq.~\eqref{eq_Bz_2D}, where the magnetic moment appears (just like with the spin example above) within the definition of the magnetic length $l_m$. For the loop radius we work with the estimate $R\sim 100\mu m$, see micrograph in Ref.~\cite{Niu_2024}.

But we have to estimate the noise differently. Here, we are dealing with a local current source, subject to current noise. For simplicity, we deploy the frequency-dependent fluctuation-dissipation theorem. For frequencies much larger than temperature, we arrive at
\begin{equation}\label{eq_SI_FDT}
    S_I(\omega)=\frac{1}{2}\int_{-\infty}^{\infty}\mathrm{d}t e^{i\omega t}\left\langle \left\{ I\left(t\right),I\left(0\right)\right\} \right\rangle \approx \frac{\hbar \omega}{Z_0}\ ,
\end{equation}
where we insert for the flux line impedance $Z_0=50\Omega$. When estimating the relaxation rate, we have to take the noise at the qubit frequency $\omega=\omega_0$.

Before analysing the exact numerical results, we first gauge our expectation with some analytical estimates. Let us position the flux line at $y_0=D/2$ and consider the expected long range behaviour of the coupling as a function of $x_0\gg D$. For the flux mode of a dc SQUID with loop area $\mathcal{A}$, we have in simplified terms,
\begin{equation}\label{eq_far_field_flux_mode}
    \frac{\phi_{1}-\phi_{2}}{l_{m}}\approx-\frac{\mathcal{A}}{x^{3}_{0}}\ .
\end{equation}
For the EMF mode, we develop the following expectation. In the absence of the superconducting structure, the irrotational gauge simply corresponds to $\mathbf{E}=-\dot{\mathbf{A}}_\text{irr}$ (up to lowest, adiabatic order in the Faraday problem), with the vector potential given as
\begin{equation}
    \mathbf{A}=\frac{\mu_{0}}{4\pi}\frac{\mathbf{r}\times\mathbf{m}}{r^{3}}\ .
\end{equation}
If we simply integrate this field over a single junction bridge of length $D$, we would get
\begin{equation}\label{eq_far_field_emf_mode}
    \frac{\phi_{1}}{l_{m}}\approx-\frac{D}{x^{2}_{0}}\ .
\end{equation}
We already observe that on this oversimplified level, the flux mode, Eq.~\eqref{eq_far_field_flux_mode}, has a much faster decay than the EMF mode, Eq.~\eqref{eq_far_field_emf_mode} ($x_0^{-3}$ versus $x_0^{-2}$). This is already a first strong indication that for an optimised placement of flux lines in a given circuit layout, it is not enough to consider the distance between the flux line and flux-tunable transmons; instead the EMF mode can impact the quality factor of \textit{all} qubits, including fixed-frequency transmons.

Note in addition, that Eq.~\eqref{eq_far_field_emf_mode} is an underestimation. We have already seen with the above example of spins, that there is a lensing effect of the electromotive field, which (as already explained) does not come from the Meissner effect (which is negligible), but from the holographic property of $\widetilde{B}_z$, Eq.~\eqref{eq_btilde}, with the extra term due to the transversal electric field at the superconducting edges. Indeed, for $\phi_1$ in Eq.~\eqref{eq_phi_2D_spin}, the asymptotic behaviour was that of $\ln(x_0)/x_0$, which is even more long-range than the estimate of Eq.~\eqref{eq_far_field_emf_mode}. But remember that Eq.~\eqref{eq_phi_2D_spin} was derived for dipole moments that are still situated between the two capacitor wings. Here, we consider devices that are outside the wings. Therefore, the long-range behaviour should transform into $x_0^{-2}$, but with a boost due to this type of lensing. And indeed, this is what we receive with the exact numeric results in Fig.~\ref{fig9_flux_lines}(b-d). In the asymptotic limit, the exact curves yield,
\begin{equation}\label{eq_far_field_emf_boosted}
    \frac{\phi_1}{l_m}\approx - \alpha\frac{D}{x_0^2}\ ,
\end{equation}
with a boosting factor $\alpha$ which, depending on the aspect ratio of the capacitor wings ($L_x/L_y$), can be as high as $\alpha\sim 10$, Fig.~\ref{fig9_flux_lines}(c). Empirically, we find that $\alpha$ roughly scales as $\alpha\sim L_x L_y/D^2$, see red lines in Fig.~\ref{fig9_flux_lines}(b-d). Consequently, the boosting seems to depend on the area occupied by the capacitor wings, consistent with the holographic property of $\widetilde{B}_z$.

The above result allows us to develop a so-called `safety distance', a minimum distance $x_0^{\text{min}}$ below which the flux line starts having a detrimental effect on the quality factor of the qubit. We find under combined use of Eqs.~\eqref{eq_Gamma_1}, ~\eqref{eq_SI_FDT}, and~\eqref{eq_far_field_emf_boosted} that
\begin{equation}
    S_{11}\approx\left(\frac{\mu_{0}\pi R^{2}}{2\Phi_{0}}\alpha\frac{D}{x^{2}_{0}}\right)^{2}\frac{\hbar\omega}{Z_{0}}\ .
\end{equation}
This provides a quality factor of the form
\begin{equation}\label{eq_quality_factor_flux_line}
    Q\approx\frac{4}{\pi^{2}\alpha^{2}}\sqrt{\frac{E_{C}}{E_{J}}}\left(\frac{x^{2}_{0}}{l_{\Delta I}D}\right)^{2}\ ,
\end{equation}
where we introduced the current noise amplitude at the qubit frequency $\Delta I=\sqrt{\hbar \omega_0^2/Z_0}$ and the magnetic length associated with it, $l_{\Delta I}=\mu_0 R^2 \Delta I/(2\Phi_0)$. Inserting a typical qubit frequency of $\omega_0=2\pi \times 5\text{GHz}$ and the aforementioned $Z_0\approx 50 \Omega$, we get $\Delta I\approx 4.4\times 10^{-8}\text{A}$. Assuming in addition $R\approx 100 \mu\text{m}$ (see above), we then get $l_{\Delta I}\approx 1.4\times 10^{-7}\text{m}$.

We now solve Eq.~\eqref{eq_quality_factor_flux_line} for $x_0^{Q}$ to find a critical (minimal) distance for a flux line placement to guarantee a given target quality factor $Q$. We get
\begin{equation}
    x_0^Q=Q^{\frac{1}{4}}l_0\ ,
\end{equation}
with the reference length $l_0=(\sqrt{E_J/E_C}\pi^2\alpha^2/4)^{1/4}\sqrt{l_{\Delta I}D}$. Using above estimate for $l_{\Delta I}$ as well as $\alpha\sim 10$ and $D\approx 20\mu\text{m}$, we get $l_0\approx 12\mu\text{m}$. If the target quality factor is $Q\approx 10^6$, then $x_0^Q\approx 380 \mu\text{m}$. However if one wants to push for a quality factor of $Q\approx 10^8$, this distance increases to $x_0^Q\approx 1200 \mu\text{m}$.

Overall, we note that upon ignoring the Faraday effect, one would completely miss the impact of flux lines on the quality factor of qubits, especially for single-junction transmons which are nominally not flux-sensitive (i.e., not sensitive to stationary flux). Instead, the placement of flux lines would have only an impact on the pure dephasing rate of tunable transmons, and the long-range behaviour would be very weak, $\sim 1/x_0^3$. Including the electromotive field on the other hand reveals a detrimental effect on the quality factor, notably decaying with a weaker power law, $1/x_0^2$. In addition, there is a lensing effect due to the capacitor wings, boosting the coupling $\phi_\nu$. For a boost by a factor of 10, we get an increase of the critical safety distance by a factor of $\sqrt{10}\sim 3.2$, due to the scaling $l_0\sim \sqrt{\alpha}$. This finding has profound consequences regarding scale-up and design of quantum hardware, potentially inducing a limit on how many physical qubits of a given target quality factor can be located within a given chip area.

\section{Discussion}

In this work, we developed a framework capable of describing the coupling between superconducting charge qubits and fluctuating magnetic fields, properly taking into account the Faraday effect. We predict a fundamental upper limit of the qubit quality factor $Q$ based on universal flux noise, even for \textit{fixed-frequency} transmons. This limit is in the best case between $Q\lesssim 10^8$ and $Q\lesssim 10^{10}$. Assuming surface spins as the main source of flux noise, we identify additional possible factors which lead to a further deterioration of this upper limit, indicating that the optimisation of qubit lifetimes in transmons might soon hit a barrier. In addition, we consider the impact of flux lines on the qubit relaxation rate and introduce the notion of a safety distance, a minimal distance between a flux line and a transmon that needs to be respected in order to retain a desired quality factor, which we expect to lead to a maximum qubit density in transmon-based quantum hardware architecture.

While based on Ref.~\cite{riwar_2022}, the here developed framework significantly expands on several aspects. We show that the electromotive field in free space is strongly affected by the tangential field flowing at the superconducting surfaces via the usual London effect. We interpret this effect as holographic because the field in free space contains the information of the flux piercing the superconducting bulk, even if the latter is not expelled via the Meissner effect. We unravel a connection between conservation of Thomas-Fermi surface charges screening the electromotive field and Gauss' law of magnetism via a conformal map. We provide generally valid closed expressions for the phase drops at Josephson junctions in the irrotational gauge (where the electromotive field is captured exclusively by the vector potential). 

Applying the formalism to transmons interacting with surface spins, we examine the coupling strength as a function of the spin position, and examine the importance of different device parts for the resulting qubit quality factor (in particular the junction bridge versus the capacitor wings). The precise impact on the quality factor depends on microscopic details, most prominently the spin correlation length. When considering flux lines located far away from the device, we predict an order of magnitude boost of the coupling strength due to the aforementioned holographic property. This in turn implies that the minimum distance at which a flux line can be positioned without detrimental effect on the quality factor increases by a little more than three-fold. 

\section*{Acknowledgements}

We acknowledge many fruitful discussions with D.~P.~DiVincenzo, P.~Bushew, V.~Mourik, and G.~Catelani. Declaration on AI involvement: the conception of the project, as well as the results, their interpretation, text writing, and figures are exclusively of human origin. AI (ChatGPT) was used for literature search and for optimisation of the numerical code (e.g., suggesting the use of fast Fourier transforms or the mixing procedure to render the iterative Meissner equation convergent).

\appendix

\section{The flux-tunable asymmetric SQUID Hamiltonian} \label{app_SQUID_transformation}

We start from the Hamiltonian of a flux-tunable, asymmetric SQUID as defined in Eq.~\eqref{eq_Hamiltonian_SQUID} in the main text. Via standard trigonometric identities, this Hamiltonian can be recast into the form
\begin{equation}
H= E_{C}\widehat{N}^{2}-E_{J,\text{loop}}\cos\left(\widehat{\phi}+\delta\right),
\end{equation}
with phase dependent loop Josephson energy and effective phase shift $\delta$, satisfying
\begin{align}
E_{J,\text{loop}}^2= & E_{J1}^2+E_{J2}^2+2E_{J1}E_{J2}\cos\left(\phi_1-\phi_2\right) \\
\tan(\delta) = & \frac{E_{J1}\sin(\phi_1)+E_{J2}\sin(\phi_2)}{E_{J1}\cos(\phi_1)+E_{J2}\cos(\phi_2)}\ .
\end{align}
Through the time-dependent unitary $U=e^{-i\delta \widehat{N}}$, we can cast the Hamiltonian into the form $\widetilde{H}=UHU^\dagger -i U\dot{U}^\dagger$ with
\begin{equation}\label{eq_H_unitary_classical}
    \widetilde{H}=E_C \widehat{N}^2+\dot{\delta}\widehat{N}-E_{J,\text{loop}}\cos(\widehat{\phi})\ .
\end{equation}
This Hamiltonian simplifies to the one given in Eq.~\eqref{eq_Hamiltonian_SQUID_trafo} for the symmetric case $E_{J1}=E_{J2}=E_{J}$.

Note that while the above linear term $\dot{\delta}\widehat{N}$ is exact for classical time-dependent driving, the same unitary transformation yields a slightly more general result if the externally applied flux is described quantum mechanically. In that case, we add operator parts to the phase drops $\phi_{1,2}\rightarrow \phi_{1,2}+\widehat{\phi}_{1,2}$, whose dynamics is given by an additional environment Hamiltonian $H\rightarrow H+H_\text{env}$. If the only dynamics of the phase drops come from the operator terms, the unitary is no longer explicitly time-dependent, eliminating the usual quantum geometric term $\sim iU\dot{U}^\dagger$. But notice that $U=e^{i\widehat{\delta}\widehat{N}}$ has the shift $\delta$ likewise promoted to an operator, which thus does not commute with $H_\text{env}$. Upon a first order expansion,
\begin{equation}
    UH_\text{env}U^\dagger \approx H_\text{env}+i[H_\text{env},\widehat{\delta}]\widehat{N}\ ,
\end{equation}
we recover the quantum equivalent of the classical result, Eq.~\eqref{eq_H_unitary_classical}, since the commutator term is simply the time-derivative of the operator $\widehat{\delta}$ in the Heisenberg picture.

\section{Relaxation and dephasing rates}\label{app_relaxation_and_dephasing}

In the transmon regime $E_{J,\text{loop}}\gg E_C$, we can project the circuit Hamiltonian onto the qubit basis $H\approx \omega_0\sigma_z/2$ with $\omega_0\approx \sqrt{2E_{J,\text{loop}}E_C}$. Including the phase fluctuations $\widehat{\phi}_{1,2}$ up to first order, we get to the open quantum system Hamiltonian (after the above unitary)
\begin{equation}
    \widetilde{H}\approx \frac{\omega_0+\widehat{D}}{2}\sigma_z+\dot{\widehat{R}}\sigma_{x}+H_\text{env}\ ,
\end{equation}
with the dephasing and relaxation mechanisms
\begin{align}
    \widehat{D}&=\widehat{\phi}_1\partial_{\phi_1}\omega_0+\widehat{\phi}_2\partial_{\phi_2}\omega_0\ ,\\ 
    \widehat{R}&=\left(\frac{E_{J,\text{loop}}}{8E_{C}}\right)^{\frac{1}{4}}\left[\widehat{\phi}_1\partial_{\phi_1}\delta+\widehat{\phi}_2\partial_{\phi_2}\delta\right]\ ,
\end{align}
and the Heisenberg equation of motion $\dot{\widehat{R}}=i[H_\text{env},\widehat{R}]$. The resulting relaxation and dephasing rates are obtained via standard Fermi's golden rule
\begin{align}
    \Gamma_1&=\int^{\infty}_{-\infty}dte^{i\omega t}\text{tr}_{\text{env}}\left[\left\{\dot{\widehat{R}}\left(t\right),\dot{\widehat{R}}\left(0\right)\right\}\rho_{\text{env}}\right]\ ,\\
    \Gamma_\varphi&=\frac{1}{4}\int^{\infty}_{-\infty}dt\,\text{tr}_{\text{env}}\left[\left\{\widehat{D}\left(t\right),\widehat{D}\left(0\right)\right\}\rho_{\text{env}}\right]\ .
\end{align}
For a symmetric SQUID $E_{J1}=E_{J2}\equiv E_{J}$, we arrive at Eqs.~\eqref{eq_Gamma_1} and~\eqref{eq_Gamma_varphi}. For a single junction, on the other hand, $E_{J1}\equiv E_J$ and $E_{J2}=0$, we find $\Gamma_\varphi=0$ whereas a nonzero relaxation rate remains, $\Gamma_1\neq 0$, see main text.

\section{Formally exact kernels for computation of $\mathbf{A}_\text{irr}$}\label{app_kernel_Airr}

In the main text, we provide formally exact expressions for the vector potential in the irrotational gauge, $\mathbf{A}_\text{irr}$, in the language of the conformal space illustrated in Fig.~\ref{fig4_conformal}. The corresponding Eq.~\eqref{eq_Airr_conformal} requires as input the kernels,
\begin{align}\nonumber
    K_{v}\left(v,w,w^{\prime}\right)&=-\frac{2}{VW}\sum_{n_{v}}\sum_{n_{w}>0}\frac{k_{n_{w}}\mathrm{e}^{ik_{n_{v}}v}}{k^{2}_{n_{v}}+k^{2}_{n_{w}}}\\&\times\sin\left(k_{n_{w}}w\right)\cos\left(k_{n_{w}}w^{\prime}\right)\\\nonumber
    K_{w}\left(v,w,w^{\prime}\right)&=-\frac{i}{VW}\sum_{n_{v}\neq0}\sum_{n_{w}}\frac{k_{n_{v}}\mathrm{e}^{ik_{n_{v}}v}}{k^{2}_{n_{v}}+k^{2}_{n_{w}}}\\&\times \cos\left(k_{n_{w}}w\right)\cos\left(k_{n_{w}}w^{\prime}\right)\ ,
\end{align}
where $k_{n_v}=2\pi n_v/V$ and $k_{n_w}=2\pi n_w/W$ and $n_{v,w}\in\mathbb{Z}$.

\begin{figure}
    \centering
    \includegraphics[width=0.7\linewidth]{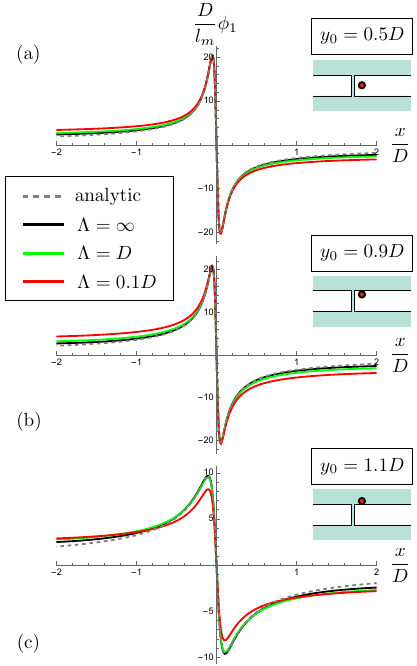}
    \caption{Comparison of the analytic result of Eq.~\eqref{eq_phi_2D_spin} (gray dashed) with numerical calculations including the Meissner effect as described in Eq.~\eqref{eq_Meissner_iteration}. A spin is placed at $x_0=0$ and $y_0=0.5D$ (a), $y_0=0.9D$ (b), as well as $y_0=1.1D$ (c). These spin positions are graphically represented in the insets (red dots). To smooth out the divergence, the spin is offset with respect to the z=0 plane by $\delta z=0.05D$. The numerical curves are calculated for $\Lambda=\infty$ (black), $\Lambda=D$ (green), and $\Lambda=0.1D$ (red).}
    \label{fig10_Meissner}
\end{figure}

\section{Demonstrating irrelevance of Meissner screening}\label{app_Meissner_irrelevant}

In the main text, we state that the Meissner effect has only a weak impact on the resulting irrotational $\phi_1$. We here illustrate this statement by comparing the analytic result obtained in the absence of Meissner screening for a magnetic dipole coupling to infinitely large capacitor wings, Eq.~\eqref{eq_phi_2D_spin}, with numerical calculations in the presence of Meissner screening. For the numerical calculation, we have to take a finite capacitor wing size, and simply choose the wings sufficiently large for finite size effects to become negligible. We implement the Meissner screening iteratively as described in Eq.~\eqref{eq_Meissner_iteration} and surrounding text.

In Fig.~\ref{fig10_Meissner}(a-c), we show $\phi_1$ as a function of the junction bridge position $x$, for a dipole placed at $x_0=0$ and $y_0=0.5D,0.9D,1.1D$ . We choose the coordinate system as in Fig.~\ref{fig6_bridge_vs_wing}, where the lower and upper capacitor wings are at $y<0$ and $y>D$, respectively. We compare Pearl lengths $\Lambda=\infty,D,0.1D$ (black, green, and red lines). With $\Lambda\approx 2\mu\text{m}$ for $\sim 25\text{nm}$ aluminium films~\cite{LopezNunez_2025} and a typical bridge length of $D\approx 20\mu\text{m}$, the last parameter choice $\Lambda=0.1D$ (red) is likely close to a realistic device. Overall, we see that the Meissner screening has a very negligible effect on $\phi_1$ even for realistic Pearl lengths.

\clearpage

\bibliography{sn-bibliography}

\end{document}